\documentclass[twocolumn,10pt,prd,nofootinbib,preprintnumbers]{revtex4-1}
\usepackage[utf8]{inputenc}

\usepackage[dvips]{graphicx}

\usepackage{epsfig,amsmath,amssymb,verbatim,mathrsfs,array,layout,textcomp,amssymb,latexsym,slashed,graphicx,booktabs,color,mathtools,tikz}
\usepackage[
    colorlinks=true,
    linkcolor=blue,
    urlcolor=blue,
    citecolor=blue
]{hyperref}
\usepackage{tikz-feynman}
\usepackage[normalem]{ulem}

\newcommand{\beq}{\begin{equation}}
\newcommand{\eeq}{\end{equation}}
\newcommand{\bea}{\begin{eqnarray}}
\newcommand{\eea}{\end{eqnarray}}

\def\beqa{\begin{eqnarray}}
\def\eeqa{\end{eqnarray}}

\newcommand{\bv}{\left(\begin{array}{c}}
\newcommand{\ev}{\end{array}\right)}
\newcommand{\bmtwo}{\left(\begin{array}{cc}}
\newcommand{\bmthree}{\left(\begin{array}{ccc}}
\newcommand{\emn}{\end{array}\right)}
\newcommand{\bmtwoc}{\left\{\begin{array}{cc}}
\newcommand{\bmthreec}{\left\{\begin{array}{ccc}}
\newcommand{\emnc}{\end{array}\right\}}
\newcommand{\ba}{\begin{array}}
\newcommand{\ea}{\end{array}}

\def\lsim{\mathrel{\rlap{\lower4pt\hbox{\hskip1pt$\sim$}}
\raise1pt\hbox{$<$}}}    
\def\gsim{\mathrel{\rlap{\lower4pt\hbox{\hskip1pt$\sim$}}
\raise1pt\hbox{$>$}}}

\begin{document}

\font\mini=cmr10 at 0.8pt

\title{ 
Phases of Dark Matter from Inverse Decays
}

\author{Ronny Frumkin${}^{1,2}$}\email{ronny.frumkin@weizmann.ac.il}
\author{Yonit Hochberg${}^{1,3}$}\email{yonit.hochberg@mail.huji.ac.il}
\author{Eric Kuflik${}^{1,3}$}\email{eric.kuflik@mail.huji.ac.il}
\author{Hitoshi Murayama${}^{4,5,6}$}\email{hitoshi@berkeley.edu, hitoshi.murayama@ipmu.jp}
\affiliation{${}^1$Racah Institute of Physics, Hebrew University of Jerusalem, Jerusalem 91904, Israel}
\affiliation{${}^2$Department of Particle Physics and Astrophysics, Weizmann Institute of Science, Rehovot 7610001,
Israel}
\affiliation{${}^3$Laboratory for Elementary Particle Physics,
 Cornell University, Ithaca, NY 14853, USA}
\affiliation{${}^4$Ernest Orlando Lawrence Berkeley National Laboratory, University of California, Berkeley, CA 94720, USA}
\affiliation{${}^5$Department of Physics, University of California, Berkeley, CA 94720, USA}
\affiliation{${}^6$Kavli Institute for the Physics and Mathematics of the
  Universe (WPI), \\University of Tokyo,
  Kashiwa 277-8583, Japan}

\begin{abstract}
 
 Inverse decays are an interesting avenue for producing dark matter in the early universe.  We study in detail various phases of dark matter parameter space where inverse decays control its abundance, expanding on our work of INDY dark matter and going beyond. The role of initial conditions and the impact of departure from kinetic equilibrium are investigated as well. We show how these inverse decay phases can arise in theories of a kinetically mixed dark photon and dark Higgs, with promising prospects for detection at upcoming experiments.  
  
\end{abstract}

\maketitle

\section{Introduction}\label{sec:intro}

The identity of dark matter (DM) remains one of the most pressing puzzles of modern day physics. Recent years have seen a surge of new ideas suggesting various processes in the early universe that could control the DM relic abundance, including SIMPs~\cite{Hochberg:2014dra,Hochberg:2014kqa}, ELDERs~\cite{Kuflik:2015isi}, forbidden channels~\cite{Griest:1990kh,DAgnolo:2015ujb}, coscattering~\cite{DAgnolo:2017dbv}, zombies~\cite{Kramer:2020sbb} and more, where in some cases decay interactions play a role in the evolution over time of the DM   abundance~\cite{Dror:2016rxc,Dror:2017gjq,Kopp:2016yji,DAgnolo:2018wcn,Fitzpatrick:2020vba,Kim:2019udq,Asadi:2021yml,Asadi:2021pwo,Dror:2016rxc,Dror:2017gjq,Berlin:2016vnh,Morrissey:2009ur,Cohen:2010kn,Bandyopadhyay:2011qm,Farina:2016llk,Kim:2019udq,Hochberg:2018vdo,Feng:2003xh,Kaplinghat:2005sy,Moroi:1999zb,Acharya:2009zt,Hall:2009bx,Berlin:2016vnh,Morrissey:2009ur,Azatov:2021ifm,Garny:2017rxs,Yaguna:2008mi,Garny:2019kua}. Among these, an exciting possibility is for inverse decays of the DM to control its relic abundance. Ref.~\cite{Frumkin:2021zng} proposed a new thermal DM candidate whose abundance is determined by the freezeout of inverse decays, dubbed `INDY' DM. In this work, we elaborate on the INDY mechanism and study in detail the various phases that arise when considering such inverse decays.

This paper is organized as follows. We begin in Section~\ref{sec:mechanism} by setting up the basic mechanism and in Section~\ref{sec:phases} we identify the  relevant theory phases. In Section~\ref{sec:NKE} we scrutinize the effects of departure from kinetic equilibrium. Section~\ref{sec:qft_model} presents a model that realizes these phases, along with its phenomenological prospects. We conclude in Section~\ref{sec:conclusions}.

\begin{figure*}[ht!]
	\centering
	\includegraphics[width=0.9\textwidth]{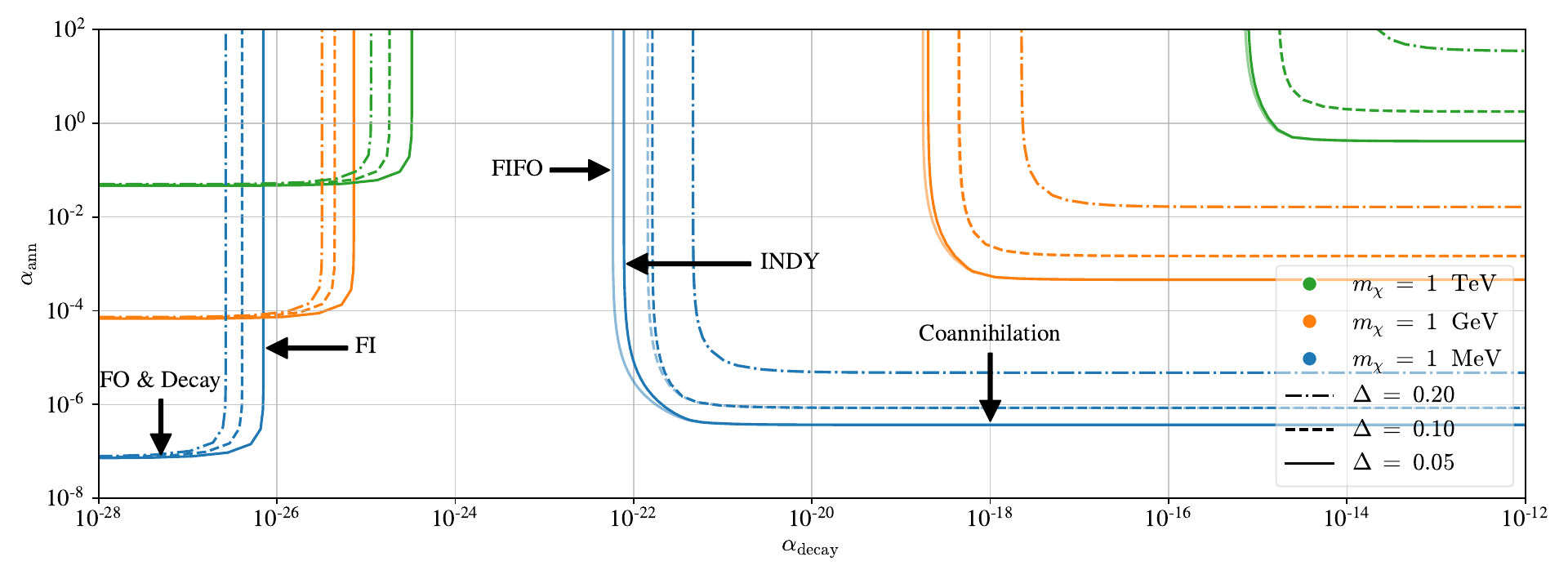}
	\caption{{\bf Phases of DM.} Annihilations versus decays phase diagram, for various mass splittings and DM masses, and different initial conditions. The solid  lines in the right (where $ \alpha_{\rm decay} >10^{-24} $) are the same as in Ref.~\cite{Frumkin:2021zng}, with initial condition of chemical equilibrium at $ x=1 $. The two phases are evident: The first is the horizontal branch,  where the relic abundance of the DM is set through the annihilations of another dark particle, with which it is in chemical equilibrium via decays and inverse decays, denoted by `coannihilation'. The second is the vertical branch, where the relic abundance of the DM is set by the freezeout of the inverse decay of the DM particle. We dub this case as `INDY' DM. The lighter-colored solid lines in the right represent the `Freeze-in-Freeze-out' (FIFO) case,  
    where the initial conditions taken are $ Y_{\chi}  = 0 $ at $ x=0 $, and the $ \chi $ abundance grows and decreases. The FIFO case coincide with INDY for most of the parameters, or otherwise very close to it. In the left side of the figure ($ \alpha_{\rm decay} >10^{-24} $) the lines represent the `Freeze-in' (FI) case, where the initial conditions taken are $ Y_{\chi}  = 0 $ at $ x=0 $, and the $ \chi $ abundance grows with time, as well as the `FO \& Decay' phase, where $\psi$  freezes out and later decays to the~DM.
    }
	\label{fig:alpha_ann_vs_alpha_decay}
\end{figure*}

\section{The Basic System}\label{sec:mechanism}

We begin by setting up the system of interest. Consider the decay and inverse decay processes $ \psi \longleftrightarrow \chi + \phi $ of a DM candidate $ \chi $ and an unstable particle $ \psi $, along with the self-annihilation process $ \psi \psi \longleftrightarrow \tilde \phi \tilde \phi\ $, with $\phi $ and $\tilde \phi$ indicating particles in equilibrium with the Standard Model (SM) bath. The relative strength between these two processes will determine which reaction controls the relic abundance of DM $\chi$ and will lead to various phases of this simple system.

The Boltzmann equations (BEs) for this system, assuming detailed balance, can be written as~\cite{Frumkin:2021zng}
\begin{eqnarray} \label{eq:Bolzmann_equations}
	\frac{dY_{\chi}}{dx} & = & - \frac{\langle\Gamma\rangle}{x H} \left( \frac{Y_{\psi}^{\rm eq}}{Y_{\chi}^{\rm eq}} Y_{\chi} - Y_{\psi} \right) \, , \\
	\frac{dY_{\psi}}{dx} & = & -\frac{dY_{\chi}}{dx}- \frac{s \langle \sigma v\rangle}{x H}\left(Y_{\psi}^{2}-Y_{\psi}^{{\rm eq}^2} \right) \, , \nonumber
\end{eqnarray} 
where $x=m_\chi/T$, $Y=n/s $ is the yield, $ s $ is the entropy density, $ H $ is the Hubble rate, $\langle \Gamma\rangle$ represents the thermally averaged (time dilation included) decay rate of $\psi \to \chi \phi$, and $\langle \sigma v\rangle$ represents the thermally averaged cross section for the annihilation $\psi \psi \to \tilde \phi \tilde \phi$. (The changes in the effective number of degrees of freedom for radiation density $ g_{*} $ and for entropy $ g_{*s} $ are taken into account by dividing $ H $ by a factor of $ \left(1 + \frac{1}{3} \frac{d{\rm ln}g_{*s}(T)}{d\ln T}\right) $;   values are taken from Ref.~\cite{Hindmarsh:2005ix}.)

We parameterize the decay width and the thermally averaged cross-section of the $ \psi $ self-annihilations by
\begin{equation} \label{eq:gamma_parameterization}
	\Gamma \equiv \alpha_{\rm decay} m_{\psi} \bar{\beta}
\end{equation} 
and
\begin{equation} \label{eq:sigma_v_parameterization}
	\langle\sigma v\rangle \equiv \frac{\alpha_{\rm ann}^{2}}{m_{\psi}^{2}}\, 
\end{equation}
respectively, where $ \alpha_{\rm decay} $ and $ \alpha_{\rm ann} $ are dimensionless couplings constants we use to parameterize the system  and $ \bar{\beta} = \sqrt{1-2\frac{\left(m_{\phi}^{2}+m_{\chi}^{2}\right)}{m_{\psi}^{2}}+\frac{\left(m_{\phi}^{2}-m_{\chi}^{2}\right)^{2}}{m_{\psi}^{4}}} $ is a phase space factor. The thermally averaged decay rate obtained by using Maxwell-Boltzmann statistics is given by 
\begin{equation}
	\left\langle \Gamma\right\rangle = \Gamma\frac{K_{1}\left(\frac{m_{\psi}}{T}\right)}{K_{2}\left(\frac{m_{\psi}}{T}\right)}\,,
\end{equation}
with $K_{1,2}$ the modified Bessel functions of the second kind.

In what follows we use 
\begin{equation}
\label{eq:delta_and_r_definition}
\Delta \equiv \frac{m_{\psi}-m_{\chi}}{m_{\chi}} \quad \text{and} \quad r \equiv \frac{m_{\phi}}{m_{\chi}}
\end{equation}
to denote the masses $ m_{\psi} $ and $ m_{\phi} $ compared to the mass of the DM $m_\chi$. We further take the number of degrees of freedom of the dark particles to be  $ g_{\chi} = g_{\psi} = 4 $ for concreteness. 
The BEs \eqref{eq:Bolzmann_equations} depend on $ m_{\phi} $ only through the decay rate $ \Gamma $; we thus take $ r = 0 $ in upcoming sections unless specified otherwise, and note that the massive $ \phi $ case ({\it i.e.} $r\neq 0$) can be inferred by augmenting the coupling $ \alpha_{\rm decay} $ of Eq.~\eqref{eq:gamma_parameterization} by the ratio of the relative allowed phase spaces.

The BEs \eqref{eq:Bolzmann_equations} admit several distinct phases that reproduce the observed DM relic abundance, depending on masses, couplings, and initial conditions.
These phases are illustrated in  Fig.~\ref{fig:alpha_ann_vs_alpha_decay}, showing the values of couplings $\alpha_{\rm decay}$ and $\alpha_{\rm ann}$ that reproduce the observed abundance of DM, $ Y_{\rm obs}$, for various cases. Throughout this paper, we parameterize the observed DM abundance as $ Y_{\rm obs} = c\; T_{\rm eq}/ m_{\chi} $, with $T_{\rm eq}= 0.8$~eV the temperature at matter-radiation equality and $ c = 0.54 $ \cite{Hochberg:2014dra}.

\section{Phases}\label{sec:phases}
 
We now delve into each of the phases that appear in Fig.~\ref{fig:alpha_ann_vs_alpha_decay}. 

\begin{figure*}[th!]
	\centering
	\includegraphics[width=1.99\columnwidth]{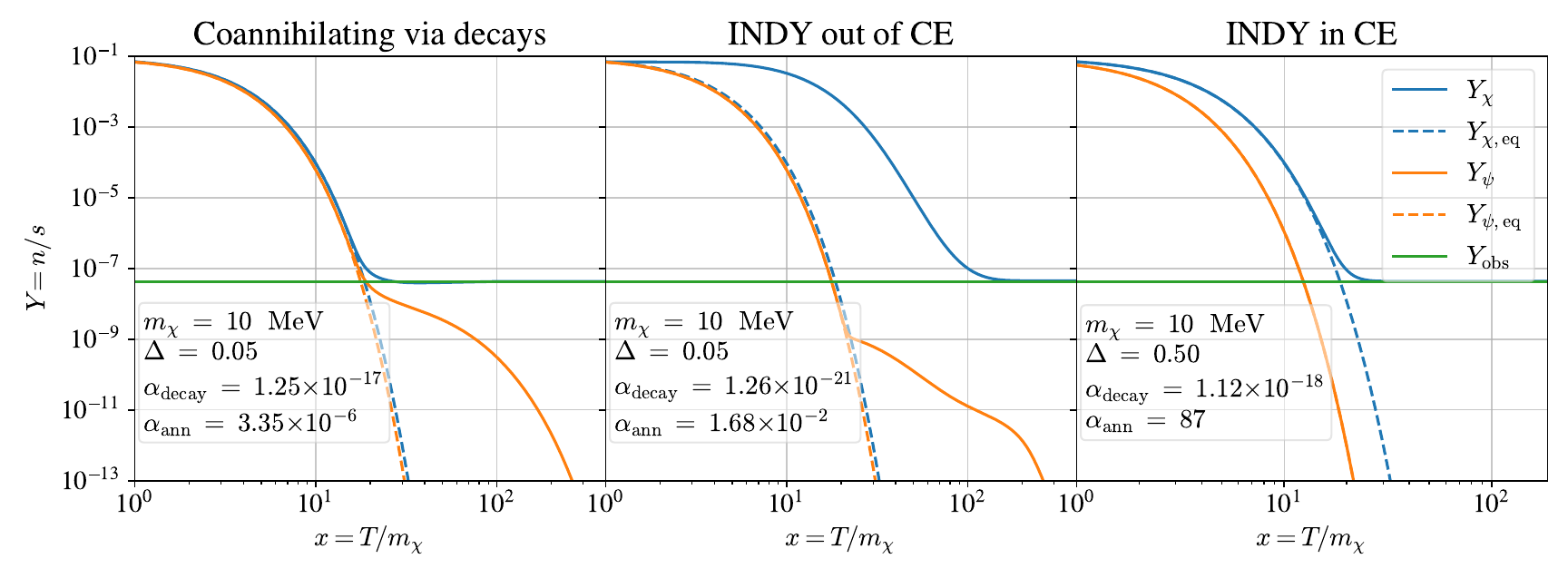}
	\caption{{\bf BE evolution.} Examples of solutions to the BEs~ \eqref{eq:Bolzmann_equations} that reproduce the observed DM relic abundance for different parameters. In each panel, we present the calculated (solid) and equilibrium (dashed) abundances for $ \chi $ (blue) and $ \psi $ (orange) as a function of $ x $, along with the observed relic abundance (green). Both panels represent the vertical branch in which the freeze out of inverse decay determines the relic abundance. The left panel represents coannihilations via decay. The middle panel corresponds to INDY DM that is out of chemical equilibrium, while the right panel corresponds to an INDY that is in chemical equilibrium.
    }
	\label{fig:examples_for_BE_solutions}
\end{figure*}

\subsection{Coannihilating via decays}\label{subsec:horizontal}

First, we consider the phase obtained for large $\alpha_{\rm decay}$, which is demonstrated by the horizontal curves of constant $\alpha_{\rm ann}$ towards the right side of Fig.~\ref{fig:alpha_ann_vs_alpha_decay}, where $\alpha_{\rm decay} >10^{-24} $.\footnote{In Fig.~\ref{fig:alpha_ann_vs_alpha_decay} we take initial conditions of equilibrium number density at $x=1$; note that this assumption does not effect this phase since the fast annihilation rate thermalizes the particles before freezeout.}
Here, the relic abundance of $\chi$ is being set by the annihilations of $\psi$ into other particles. 
The rapid decay and inverse decay rates, due to the large coupling $ \alpha_{\rm decay} $, keep the particles $\chi$ and $\psi$ in chemical equilibrium (CE) with each other, balancing their number density ratios to be as in equilibrium $ n_{\chi}/n_{\psi} = n_{\chi}^{\rm eq}/n_{\psi}^{\rm eq} $~\cite{Griest:1990kh}.
This enables one to write the system as a single BE for the sum of the number densities, 
\begin{equation}
	\label{eq:BE_horizontal_branch}
	\frac{dn}{dt}+3Hn=-\frac{\langle\sigma v\rangle}{\left(1+\frac{n_{\chi}^{\rm eq} } {n_{\psi}^{\rm eq}}\right)^{2}}\left(n^{2}-(n^{{\rm eq}})^2\right)\, ,
\end{equation}
with $ n \equiv n_{\psi} + n_{\chi} $. Eq.~\eqref{eq:BE_horizontal_branch} is general for any strongly coupled two-particle system and is not unique for inverse decays. Indeed, a similar equation is found for the well-studied case of coannhilations~\cite{Griest:1990kh}; the difference is that here the CE between the two particles is maintained via decays and inverse decays, rather than annihilations.  

The instantaneous freezeout approximation provides an estimated solution for the requisite cross section explaining the observed DM abundance, or said another way, provides a relationship between the mass $ m_{\chi} $ and the annihilation coupling $ \alpha_{\rm ann} $: 
\begin{equation} \label{eq:mvsann}
	m_\chi \simeq  \alpha_{\rm ann}^{\frac{2}{2+\Delta}}\left( m_{\rm pl} T_{\rm eq}^{1+\Delta} \right)^{\frac{1}{2+\Delta}} \, ,
\end{equation}
where ${m_{\rm pl} = 2.4\times 10^{18}}$~GeV is the reduced Planck mass~\cite{Griest:1990kh,Frumkin:2022ror}. 
The left panel of Fig.~\ref{fig:examples_for_BE_solutions} shows an example of the evolution of $ Y_{\chi} $ and $ Y_{\psi} $ over time along  this phase. Here both particles are in equilibrium before freezeout occurs, since $ \psi $ is coupled to the SM via annihilation and $ \chi $ via strong coupling to $ \psi $, and both particles depart from equilibrium simultaneously when the effective annihilation reaction rate is comparable to the Hubble rate, $ n \langle\sigma v\rangle/\left(1+{n_{\chi}^{\rm eq} }/ {n_{\psi}^{\rm eq}}\right)^{2} \sim H $. 

The power law relation of Eq.~\eqref{eq:mvsann} 
agrees well with numerical results, as is shown in the left panel of Fig.~\ref{fig:alpha_ann_vs_mass}.
The right panel of Fig.~\ref{fig:alpha_ann_vs_mass} shows the mass splitting $ \Delta $ versus the DM mass $ m_{\chi} $ at fixed values of $\alpha_{\rm ann}$ that reproduce the observed DM yield. $ \alpha_{\rm ann} $ is generally constrained by unitarity arguments, which in turn places an upper limit on the possible DM masses along this phase. 

\subsection{INDY DM} \label{subsec:vertical}

\begin{figure*}[t]
	\centering
	\includegraphics[width=\columnwidth]{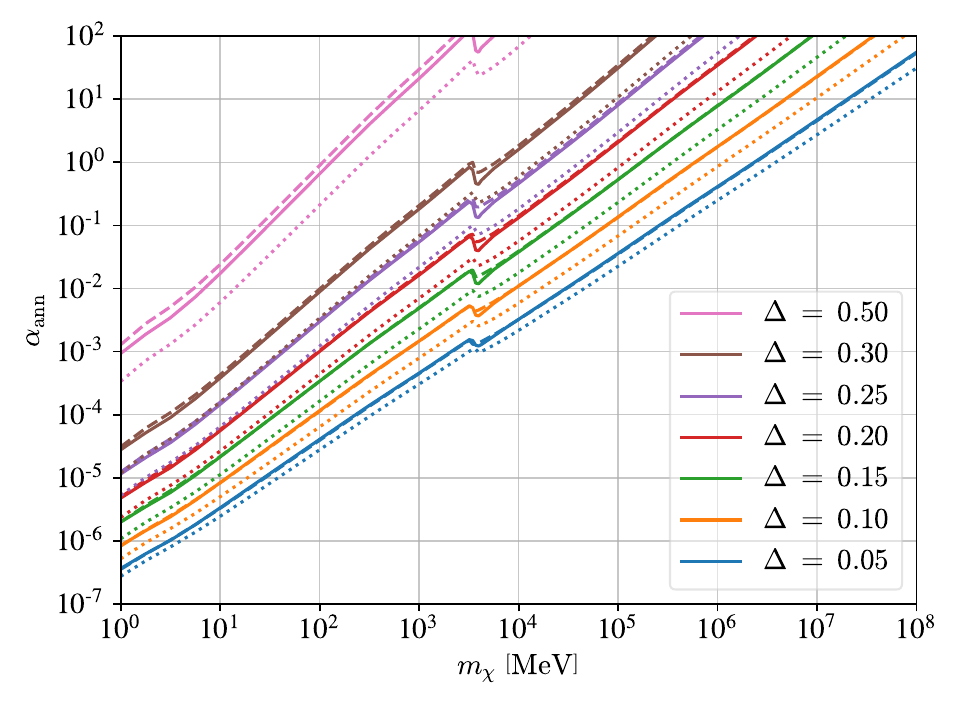}
	\includegraphics[width=\columnwidth]{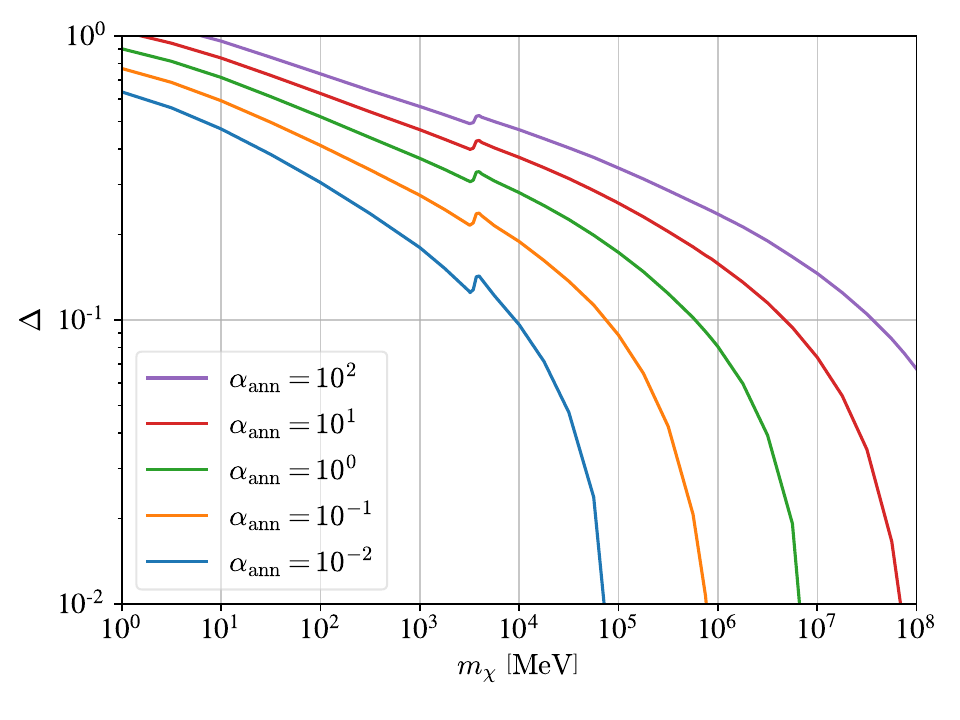}
	\caption{{\bf Coannihilating via decays.} {\it Left:} The annihilation coupling that reproduces the observed DM abundance along the horizontal phase as a function of DM mass, for various mass splittings. We show the numerical solution ({\it solid}), the instantaneous freezeout  \cite{Kolb:1990vq} ({\it dotted}) and Griest-Seckel \cite{Griest:1990kh} ({\it dashed}) approximations. {\it Right:} The mass splitting that reproduces the observed DM abundance along the coannihilation-via-decay phase as a function of DM mass, for fixed annihilation couplings. It is evident that increasing the mass or the mass splitting increases the required annihilation strength.}
	\label{fig:alpha_ann_vs_mass}
\end{figure*}

We now move to the phase obtained for large $\alpha_{\rm ann} $ values, demonstrated by the vertical curves of constant $\alpha_{\rm decay} >10^{-24} $ towards  the right side of Fig.~\ref{fig:alpha_ann_vs_alpha_decay}. 
In this phase, first proposed in Ref.~\cite{Frumkin:2021zng}, the relic abundance of $\chi$ is set by the freeze out of inverse decays.  We dub this `INverse DecaY' (INDY) dark matter.\footnote{Note that the right panel of Fig.~\ref{fig:alpha_ann_vs_mass} sets an upper bound on the mass allowed for INDY DM as well, since the INDY phase requires an annihilation coupling larger than the coannihilation phase.
Such mass limit can be altered by the use of different mechanisms for the $ \psi $ dilution; one such example is a chain of inverse decays of dark sector particles, as recently introduced in Ref.~\cite{Frumkin:2022ror}.
} 

For large values of $\alpha_{\rm ann}$ (compared to the values obtained in the coannihilating-via-decays phase), $ \psi $   preserves CE throughout the relevant times,\footnote{Or $\psi$ is much closer to CE than $ \chi $ ($ n_{\chi}/n_{\psi} \gg n_{\chi}^{\rm eq}/n_{\psi}^{\rm eq} $).} and the BEs for the yield can be approximated by a simple linear differential equation,
\begin{equation} \label{eq:BE_vertical_branch}
	\frac{dY_{\chi}}{dx}+\mathbf{a}(x)Y_{\chi}=\mathbf{b}(x)\, ,
\end{equation}
with
\begin{equation} \label{vertical_branch_a_b_definitinos}
	\mathbf{a}(x) \equiv \frac{\langle\Gamma\rangle}{x H }\frac{Y_{\psi}^{\rm eq}}{Y_{\chi}^{\rm eq}},\quad \mathbf{b}(x) \equiv \frac{\langle\Gamma\rangle}{x H }Y_{\psi}^{\rm eq}
\end{equation}
denoting the (normalized) inverse decay and decay rates for $ Y_{\chi} $, respectively.

Eq.~\eqref{eq:BE_vertical_branch} can be integrated into the form
\begin{equation}  \label{eq:vertical_branch_Y_integration}
	Y_{\chi}(x) = e^{-\mathbf{A}(x)} Y_{\chi}\left(1\right) + \intop_{1}^{x}e^{\mathbf{A}(\eta)-\mathbf{A}(x)}\mathbf{b}(\eta)d\eta \, ,
\end{equation} 
where the initial condition used throughout this section is an equilirubium yield at $T=m_\chi$, namely $ Y_{\chi}\left(1\right) = Y_{\chi}^{\rm eq}\left(1\right) $ and 
\begin{equation} \label{vertical_branch_a_b_definitinos_2}
	\mathbf{A}(x) \equiv \intop_{1}^{x}\mathbf{a}(\xi) d\xi \, .
\end{equation}
The relic abundance is obtained by taking the limit ${ x \rightarrow \infty} $ in Eq. \eqref{eq:vertical_branch_Y_integration}, producing
\begin{equation} \label{eq:vertical_branch_Y_inf}
	Y_{\chi,\infty} = 
	e^{-\mathbf{A}_{\infty}} Y_{\chi}\left(1\right) + \intop_{1}^{\infty}e^{\mathbf{A}(\eta)-\mathbf{A}_{\infty}}\mathbf{b}(\eta)d\eta \, ,
\end{equation}
where $ \mathbf{A}_{\infty} \equiv \mathbf{A} (\infty) $.

Eq.~\eqref{eq:vertical_branch_Y_inf} can be understood intuitively.  For a short interval $ dx $, the probability of a $ \chi $ particle to undergo inverse decay is $ \mathbf{a}(x)dx $. Therefore, the probability for particles at $ x=1 $ not to inverse decay at all is given by $ e^{-\intop_{1}^{\infty}\mathbf{a}\left(\xi\right)d\xi} $.
Decays of $ \psi $ are another source of $ \chi $ particles. For a short interval $ d\eta $ at the time $ \eta $, the number of $ \chi $ particles created by $\psi$ decays is $ \mathbf{b}(\eta) d\eta $, and the probability of each not to inverse decay again is given by $ e^{-\intop_{\eta}^{\infty}\mathbf{a}\left(\xi\right)d\xi} $. The total contribution of the decays is given by summing over all relevant times from $ \eta=1 $ to $ \eta=\infty $, which explains Eq.~\eqref{eq:vertical_branch_Y_inf}.

The two sources of $ \chi $ particles---from early times $ x<1 $ (through initial conditions) and from $ \psi $ decays at later times $ x>1 $---yield two types of INDY dark matter.  When the contribution from the former dominates the relic abundance, namely the $Y_{\chi}\left(1\right) $ term in Eq. \eqref{eq:vertical_branch_Y_inf} is dominant, the INDY DM candidate is out of CE. When the contributions from the subsequent $ \psi $ decays are most relevant,  the second $ \mathbf{b}(x) $ term in Eq.~\eqref{eq:vertical_branch_Y_inf} dominates the relic, and the INDY DM candidate follows CE. 
For a given DM mass and splitting, we determine numerically what kind of INDY DM arises by examining the ratio $w$,
\begin{equation} \label{eq:w_factor_definition}
	w \equiv \frac{\intop_{1}^{\infty}e^{\mathbf{A}(\eta)}\mathbf{b}(\eta)d\eta}{Y_{\chi}^{\rm eq}\left(1 \right)}\,,
\end{equation} 
and establishing whether $w>1$ (in CE) or $w<1$ (out of CE).

\begin{figure*}[t]
	\centering
	\includegraphics[width=\columnwidth]{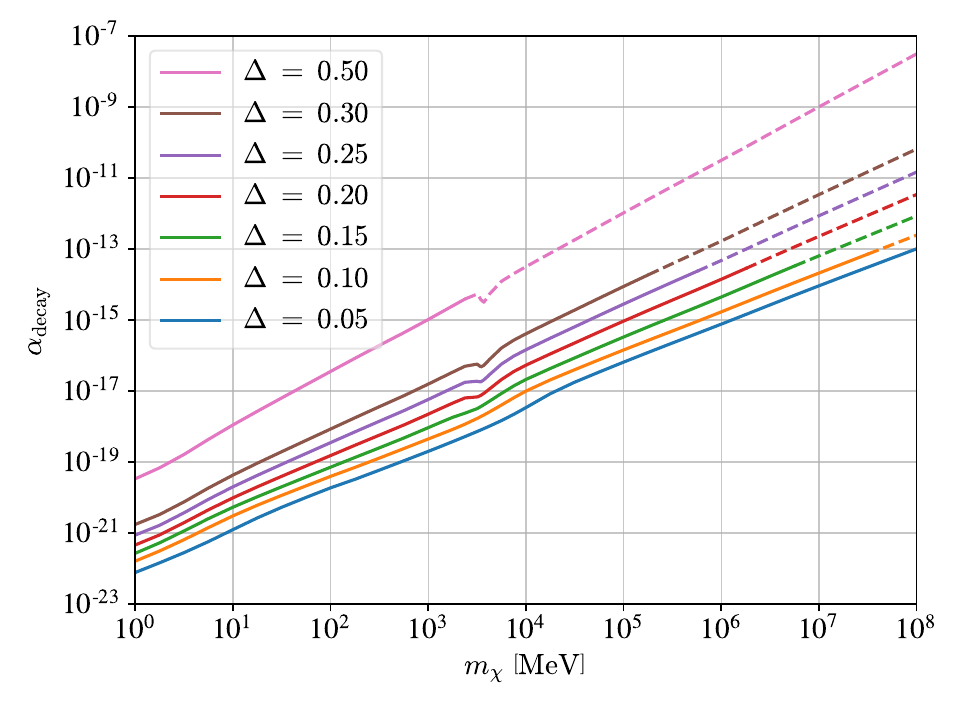}
	\includegraphics[width=\columnwidth]{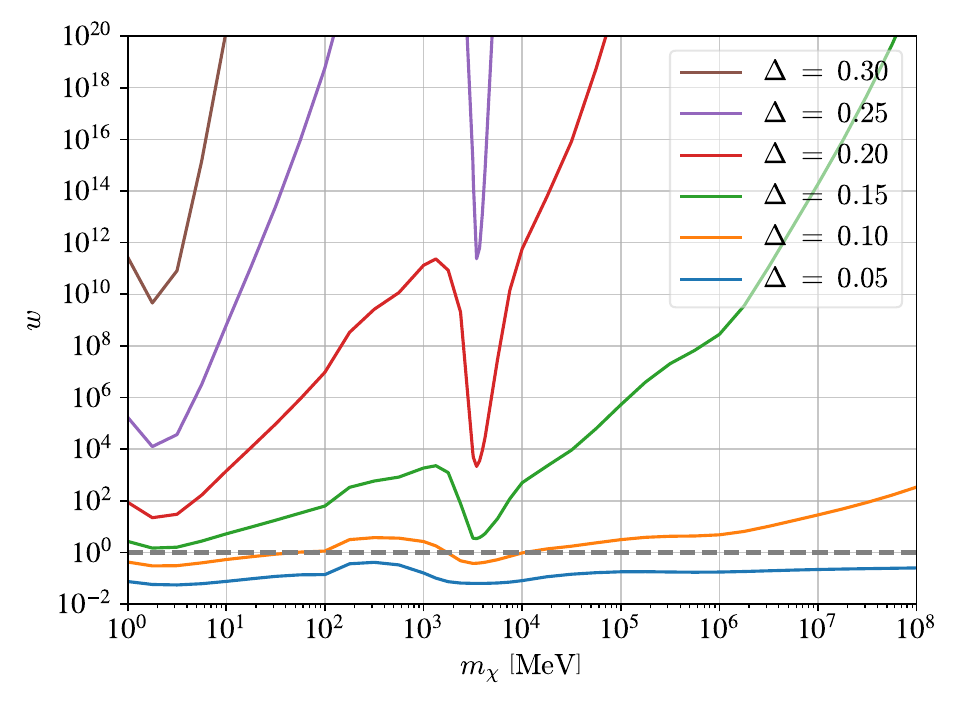}
	\caption{{\bf INDY DM.} {\it Left:} The decay coupling that reproduces the observed DM relic abundance along the vertical branches (for $\alpha_{\rm decay}>10^{-24}$) as a function of DM mass, for various mass splittings. We show the numerical solution to the Boltzmann equations; the dashed region of the curves indicates where the effective coupling $\alpha_{\rm ann}$ that we use to parameterize the $\psi$ annihilation cross section exceeds $100$ along the horizontal phase. {\it Right:} The ratio $w$ in Eq.~\eqref{eq:w_factor_definition} as a function of DM mass, for fixed mass splitting, along the solutions of INDY DM shown in the left panel. The dashed gray curve indicates $w=1$, where $w>1$ (or $w<1$) indicates where $\psi$ decays (or initial conditions) dominate the $\chi$ abundance.
    } 
	\label{fig:alpha_decay_as_function_of_m_chi_and_w}
\end{figure*}

The left panel of Fig.~\ref{fig:alpha_decay_as_function_of_m_chi_and_w} shows $ \alpha_{\rm decay} $ as a function of $  m_{\chi} $ for various mass splittings  $ \Delta $, where the observed abundance is obtained and DM is an INDY. The right panel of Fig.~\ref{fig:alpha_decay_as_function_of_m_chi_and_w} depicts $w$ along the DM solutions, indicating that for small mass splittings ({\it e.g.} $ \Delta = 0.05 $) the initial $ \chi $ particles dominate the relic abundance, while for large mass splittings ({\it e.g.} $ \Delta = 0.25 $) the later $ \psi $ decays are more significant contributors to the $\chi$ abundance. 

We can understand the slope and shape of the curves in the left panel of Fig.~\ref{fig:alpha_decay_as_function_of_m_chi_and_w} of the INDY DM solutions by examining when the INDYs are either in or out of chemical equilibrium, as we now discuss.

\subsubsection{INDYs out of chemical equilibrium ($w<1$)}

Here $ \psi $ decays can be neglected, and the relic abundance of $\chi$ is determined by the $ Y_{\chi}\left(1\right) $ term in Eq.~\eqref{eq:vertical_branch_Y_inf}.  $ Y_{\chi} $ departs from chemical equilibrium early on, and the reaction rate is maximal at $ x_{*} = \Delta^{-1}$. This can be seen by approximating $ \mathbf{a} (x) $ in the non-relativistic (NR) limit as
\begin{equation} \label{eq:a_of_x_NR}
	\mathbf{a} (x) \approx \frac{3}{\pi}\sqrt{\frac{10}{g_{*}}}\frac{g_{\psi}}{g_{\chi}}\bar{\beta}\left(1+\Delta\right)^{5/2}\frac{\alpha_{\rm decay}m_{\rm pl}}{m_{\chi}}x e^{-\Delta x}\, ,
\end{equation}
which is small at $ x = 1 $, and so the reaction rate is not fast enough to bring $ \chi $ toward equilibrium. 
A visual realization of such behavior is shown in the middle panel of Fig.~\ref{fig:examples_for_BE_solutions}.

The approximated relic abundance in this case is given by $ Y_{\chi,\infty} \approx e^{-\mathbf{A}_{\infty}} Y_{\chi}^{\rm eq}\left( 1 \right) $, where $ \mathbf{A}_{\infty} $ is obtained by integrating Eq.~\eqref{eq:a_of_x_NR},
\begin{equation} \label{eq:A_infinity_NR}
	\mathbf{A}_{\infty} \approx \frac{3}{\pi} \sqrt{\frac{10}{g_{*}}} \frac{g_{\psi}}{g_{\chi}} \bar{\beta}  \frac{ \left(1+\Delta \right)^{7/2}e^{-\Delta }}{ \Delta^{2} } \frac{\alpha_{\rm decay}m_{\rm pl}}{m_{\chi}} \, .
\end{equation}
Since the logarithm of the ratio between the initial and observed abundance $ \mathbf{A}_{\infty} \approx \ln \left( \frac{ Y_{\chi}^{\rm eq} \left( 1 \right) }{ Y_{\rm obs} } \right) $ is of order unity, using Eq.~\eqref{eq:A_infinity_NR} we find that the decay coupling which  reproduces the observed abundance scales as
\begin{equation} \label{eq:alpha_decay_out_of_CE_INDY}
	\alpha_{\rm decay} \sim \frac{m_{\chi}}{m_{\rm pl}} \, .
\end{equation}
This linear dependence between the $ \alpha_{\rm decay} $ and $ m_{\chi} $ explains the linear shape and slope of the small splitting $ \Delta =0.05,\;0.1 $ curves in the left panel of Fig.~\ref{fig:alpha_decay_as_function_of_m_chi_and_w}. The dependence on $ \Delta $ comes only through the unwritten  prefactor in Eq.~\eqref{eq:alpha_decay_out_of_CE_INDY} and not through the exponent of a mass scale as in Eq.~\eqref{eq:mvsann}. Note that  since the relation in Eq.~\eqref{eq:alpha_decay_out_of_CE_INDY} is independent of $ T_{\rm eq} $, the couplings that reproduce the DM relic abundance for this case are much smaller than in other mechanisms (compare {\it e.g.} Eq.~\eqref{eq:mvsann}).

A comment is in order regarding initial conditions. Annihilation mechanisms ({\it e.g.} WIMP, SIMP and co-annihilation) are generally independent of initial conditions, since their reaction rate decreases with time up until freezeout occurs. However, for inverse decays initial conditions may play a role, since the reaction rate does not monotonically decrease. 
The potential initial condition dependence in this case is also manifested by the explicit dependence on  $ Y_{\chi} \left( 1 \right) $ in Eq.~\eqref{eq:vertical_branch_Y_inf} for INDY dark matter. 

\begin{figure*}[th!]
	\centering
	\includegraphics[width=\columnwidth]{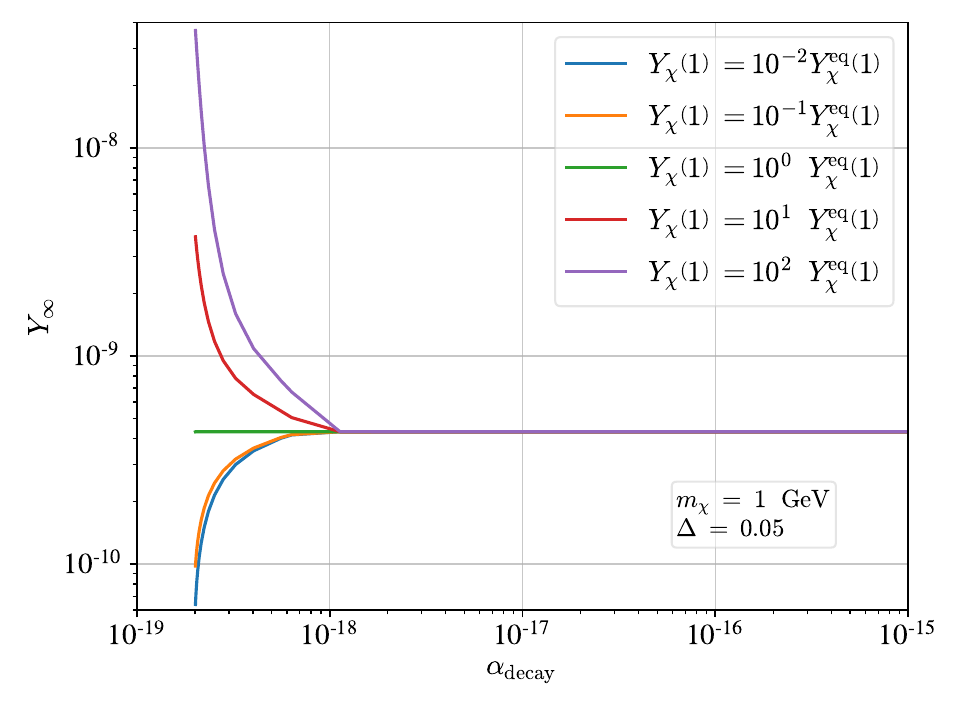}
	\includegraphics[width=\columnwidth]{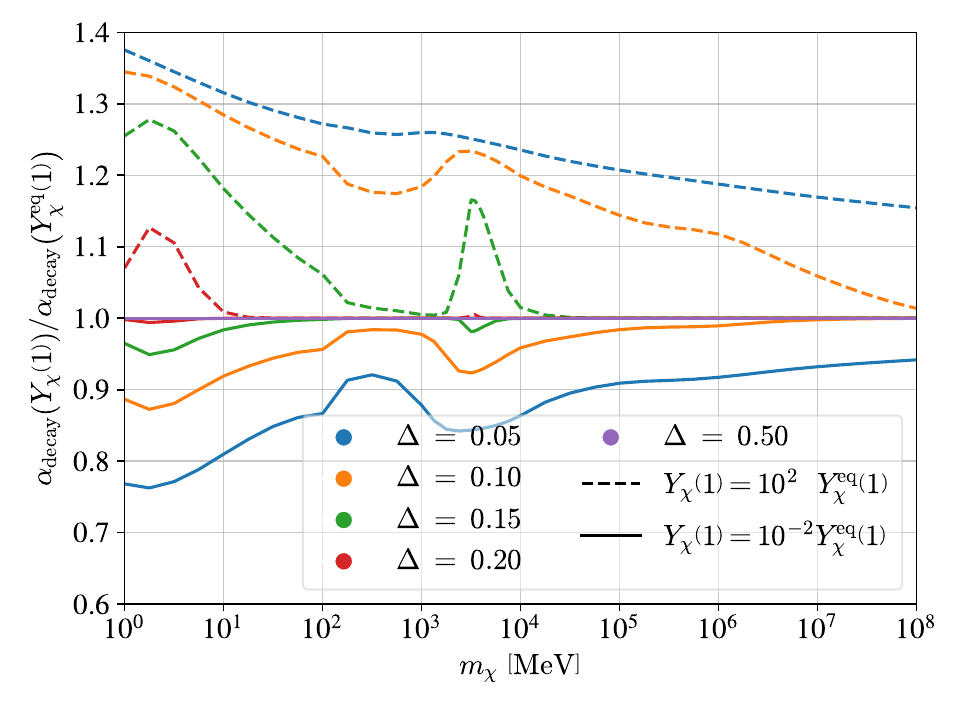}
	\caption{{\bf INDY DM.} {\it Left:} The relic abundance obtained versus the couplings presented in Fig.~\ref{fig:alpha_ann_vs_alpha_decay} for $ m_{\chi} = 1 \ {\rm GeV}$ and $\Delta = 0.05 $, with different initial condition values. {\it Right:} The ratio of the decay couplings obtained for various ratios of initial conditions to equilibrium initial conditions, as a function of DM mass. }
	\label{fig:different_initial_conditions}
\end{figure*}

The left panel of Fig.~\ref{fig:different_initial_conditions} demonstrates the above discussion. The colored curves show the relic abundance for a fixed DM mass $m_\chi= 1$~GeV and splitting $\Delta=0.05$, corresponding to the solid orange curve in Fig.~\ref{fig:alpha_ann_vs_alpha_decay}, as we vary the yield at $x=1$ from its equilibrium value. 
The larger values of $\alpha_{\rm decay}$ correspond to the coannihilating-via-decays phase, where the annihilation of $\psi$'s controls the $\chi$ abundance, and as expected is insensitive to the initial conditions. In contrast, for the smaller values of $\alpha_{\rm decay}$ that correspond to INDY DM, we see that the initial conditions can potentially impact the DM relic abundance. 

The right panel of Fig.~\ref{fig:different_initial_conditions} shows the impact that the change in initial conditions has on the value of the coupling $\alpha_{\rm decay}$ along INDY DM. As is evident, while the relic abundance might vary as the initial conditions are changed, the change to the coupling required to reproduce the observed relic abundance is minimal. The reason can be traced to the exponential dependence on $ \alpha_{\rm decay} $ when the INDY is  out of chemical equilibrium, as discussed earlier: large changes in the prefactor can be compensated by small relative changes in the exponent, namely small changes to the coupling (see Eq.~\eqref{eq:vertical_branch_Y_inf}).

We learn that the impact of initial conditions on the coupling $ \alpha_{\rm decay} $ for INDY DM is at worst logarithmic in the ratio of the initial abundance to the equilibrium one.

\subsubsection{INDYs in chemical equilbrium ($w>1$)}
For large values of $ m_{\chi} $ and $ \Delta $, the coupling $ \alpha_{\rm decay} $  
is large enough such that the inverse decay rate $ \mathbf{a}(x) $ keeps the abundance $ Y_{\chi} $ close to its equilibrium value at $ x \sim 1 $. This can be seen for example in the right panel of Fig.~\ref{fig:examples_for_BE_solutions}. 
Thus, the vast majority of the initial $ \chi $ particles at $ x = 1 $ are removed, and the relic abundance is controlled by the $ \psi $ decay term in Eq.~\eqref{eq:vertical_branch_Y_inf}.

In the NR limit,
\begin{equation} \label{eq:b_in_NR_approxmation}
	\mathbf{b} (x) = \frac{3^{3} 5^{3/2}}{2^{2}\pi^{9/2}} \frac{ g_{\psi} \bar{\beta} \left(1+\Delta\right)^{5/2} }{ g_{*s}\sqrt{g_{*}} }  \frac{\alpha_{\rm decay}m_{\rm pl}}{m_{\chi}} x^{5/2}e^{-\left(1+\Delta\right)x} \, .
\end{equation}
By using the saddle point approximation, as detailed in Appendix~\ref{appx:sec:saddle_point}, one can show that the relic abundance of DM scales with DM mass and decay coupling (up to logarithmic corrections) as
\begin{equation} \label{eq:Y_inf_indy_in_CE}
	Y_{\chi,\infty} \sim \left(\frac{\alpha_{\rm decay}m_{\rm pl}}{m_{\chi}}\right)^{-\frac{1}{\Delta}} \, .
\end{equation}

Matching the relic abundance in Eq. \eqref{eq:Y_inf_indy_in_CE} with the observed value yields the relation 
\begin{equation} \label{eq:alpha_in_CE_INDY}
	\alpha_{\rm decay} \sim \frac{ m_{\chi}^{1+\Delta} }{ T_{\rm eq}^{\Delta} m_{\rm pl} } \, .
\end{equation}
This explains the linear shape and slope of the larger mass splitting $ \Delta \ge 0.1 $ curves in Fig.~\ref{fig:alpha_decay_as_function_of_m_chi_and_w}. In INDY dark matter, even for larger DM masses one obtains small decay couplings, significantly smaller than couplings for the WIMP.

\subsection{FI and FIFO} \label{subsec:FI_vertical}

As shown in Ref.~\cite{Hall:2009bx}, there is theoretical motivation to consider DM with a negligibly small abundance at very early times. 
This case, which corresponds to initial conditions of $ Y_{\chi}\left(0\right)=0 $, leads to two more possible classes of solutions of the BEs~\eqref{eq:Bolzmann_equations}, as indicated by the vertical curves in Fig.~\ref{fig:alpha_ann_vs_alpha_decay} which are not the INDY phase.

\begin{figure*}[t]
	\centering
	\includegraphics[width=\columnwidth]{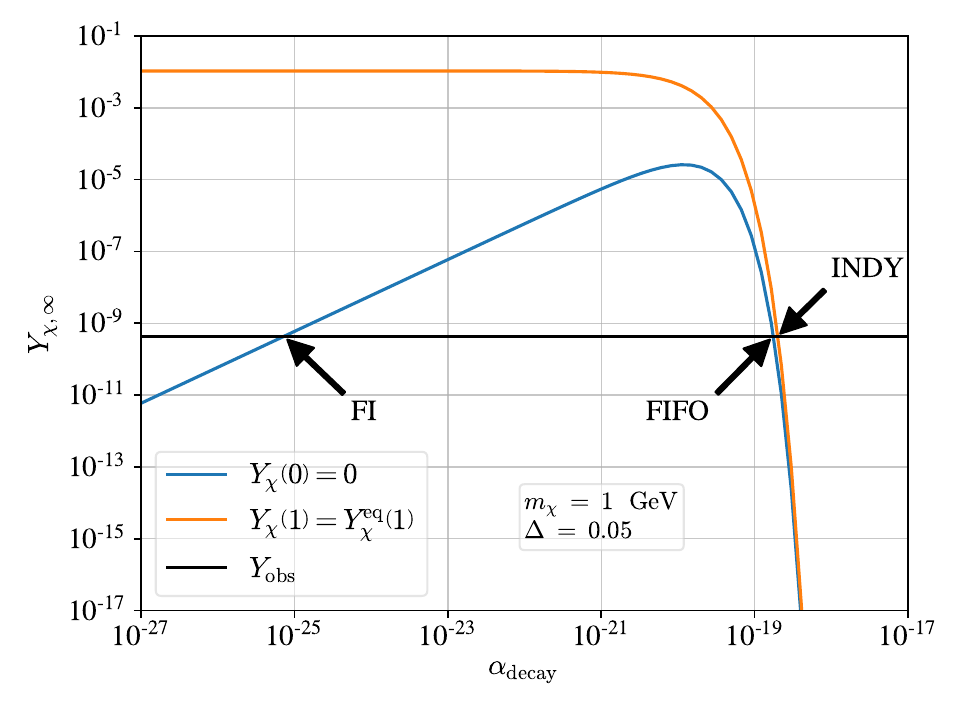}
	\includegraphics[width=\columnwidth]{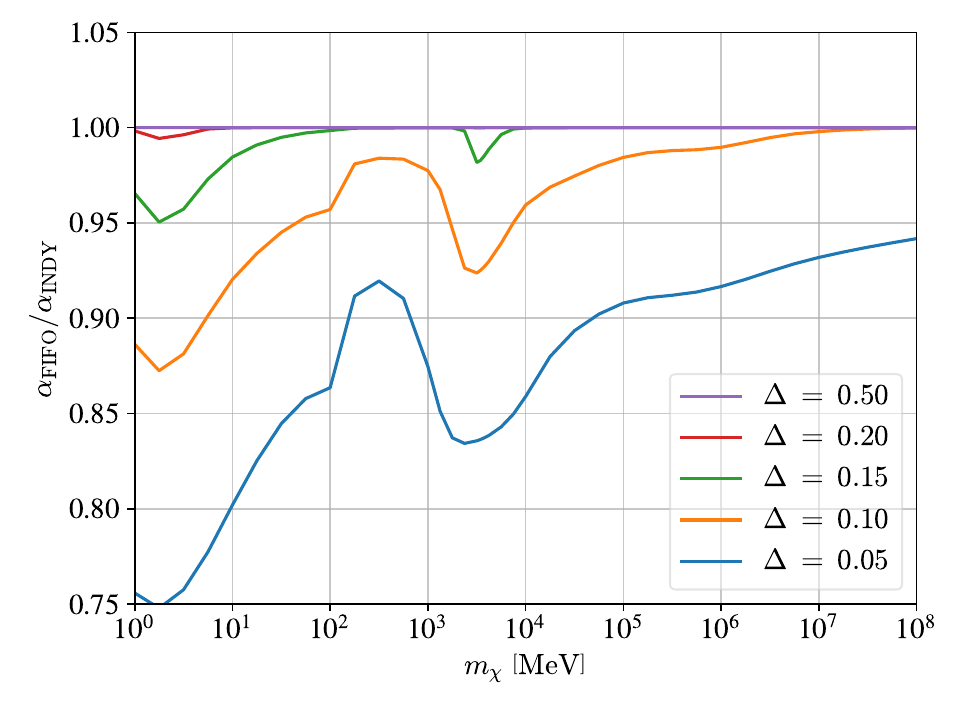}
	\caption{{\bf FIFO, FI and INDY.} {\it Left:} The relic abundance produced versus the decay coupling for different initial conditions for $ Y_{\psi} = Y_{\psi,{\rm eq}} $. The intersections between the calculated curves (blue and orange) and the observed abundance line (black) provide the relevant couplings for the FI, FIFO and INDY cases. {\it Right:} The ratio between the decay couplings, which reproduce the observed abundance for FIFO and INDY cases, as a function of DM mass, for different mass splittings. }
	\label{fig:alpha_FIFO_div_INDY}
\end{figure*}

The left panel of Fig.~\ref{fig:alpha_FIFO_div_INDY} presents the relic abundance obtained for different $\alpha_{\rm decay}$, in the limit of large annihilation values (\textit{i.e} $\psi$ is in CE), for initial conditions of $ Y_{\chi}\left(0\right)=0 $ (blue). As a reference, the INDY case, $ Y_{\chi}\left(1\right)= Y_{\chi}^{\rm eq}\left(1\right) $, is plotted as well (orange). As can be seen, there are two  possibilities  of $\alpha_{\rm decay}$ values which reproduce the relic abundance, beyond the INDY solution. The first, indicated by the crossing of the black and blue curves at small $\alpha_{\rm decay}$, is a freeze-in (FI) mechanism~\cite{Hall:2009bx}. The second, indicated by the crossing of the black and blue curves at larger values of $\alpha_{\rm decay}$, presents the case of the DM abundance growing at first and then decreasing---for some masses never tracking the equilibrium abundance. This is a freeze-in and freeze-out (FIFO) scenario. 
Examples of the evolution over time of the DM abundance for the cases of FI and FIFO DM are presented in Ref.~\cite{Frumkin:2021zng}.

For FI or FIFO to occur for a given DM mass and mass splitting, there must exist a value of $ \alpha_{\rm decay} $ which is large enough to create a sufficient number of $ \chi $ particles (through $\psi$ decays), but small enough so that inverse decays do not eliminate all DM particles and the observed abundance remains. To check when these conditions can be met in the high $\alpha_{\rm ann}$ limit, one can examine the maximal value of $ Y_{\chi,\infty} $ as a function of $ \alpha_{\rm decay} $, with initial conditions of $ Y_{\chi}\left(0\right)=0 $. By taking the relic abundance obtained in this case,
\begin{equation} \label{eq:Y_end_FI}
	Y_{\chi,\infty}=\intop_{0}^{\infty}e^{-\intop_{\eta}^{\infty}\mathbf{a}\left(\xi\right)d\xi}\mathbf{b}(\eta)d\eta \, ,
\end{equation}
matching its derivative with respect to $ \alpha_{\rm decay} $ to zero,
\begin{equation} \label{eq:Y_end_FI_derivative_alpha_decay}
	\intop_{0}^{\infty}\left[e^{-\intop_{\eta}^{\infty}\mathbf{a}\left(\xi\right)d\xi}\mathbf{b}(\eta)\left(1-\intop_{\eta}^{\infty}\mathbf{a}\left(\xi\right)d\xi\right)\right] d\eta=0\, 
\end{equation}
solving for $\alpha_{\rm decay}$ and comparing the value of the relic abundance obtained with this $ \alpha_{\rm decay} $  to $ Y_{\rm obs} $, it is possible to determine if FI or FIFO can occur for each choice of mass and mass splitting. Visually, this condition is equivalent to requiring that the blue and black curves in Fig.~\ref{fig:alpha_FIFO_div_INDY}  intersect. Numerically we find that the condition for either FI or FIFO is not fulfilled when $ \Delta < 0.1 $ and $ m_{\chi} \sim  {\rm keV} $.

A comparison between the decay couplings yielding the FIFO case $ \alpha_{\rm FIFO} $ and those yielding INDY DM $ \alpha_{\rm INDY}$ is shown in the right panel of Fig.~\ref{fig:alpha_FIFO_div_INDY}. As expected, the ratio is close to unity, in agreement with the discussion of the previous section that the coupling for freeze out has only a small dependence on initial conditions.  The dips in the curves at ${\cal O}({\rm MeV})$ and $ 100~{\rm MeV}$ to GeV DM masses correspond to variations in $g_{*}$, dominantly from electron-positron annihilation and the QCD phase transition, respectively.
On the other hand, the decays couplings for the FI case are almost independent of the mass scale, as in Ref.~\cite{Hall:2009bx}. This is because for small couplings, as occurs in FI, the inverse decays can be neglected, and the relic abundance in Eq.~\eqref{eq:Y_end_FI} is approximated to
\begin{equation}
	Y_{\chi,\infty} \approx \intop_{0}^{\infty}\mathbf{b}(\eta)d\eta \, .
\end{equation}
Since both $ Y_{\rm obs} $ and $ \mathbf{b}(\eta) $ are explicitly inversely proportional to $ m_{\chi} $, the coupling values depend on the mass scale only via $ g_{*} $ and $ g_{*s} $.

\subsection{Freezeout and Decay} \label{sec:initial_contions_general_phase_diagram}

The last phase presented in Fig.~\ref{fig:alpha_ann_vs_alpha_decay}, which is manifested by the horizontal curves towards the left, is obtained by extremely low $ \alpha_{\rm decay} $ values, and initial conditions of  $ Y_{\chi}\left(0\right)=0 $. 
In this case, the $ \psi $ annihilation is entirely identical to the WIMP, and the abundance of $\chi$ before and during the $\psi$ freezeout is negligible. Once  $ \psi $ freezes out, it decays into $ \chi $, providing a similar relic abundance to the corresponding WIMP case, up to a factor of $ \left( 1 + \Delta \right)^{-1} $.

Note that the monotonic increase  of $ \alpha_{\rm decay} $ with $ \alpha_{\rm ann} $ for the curves in the left side of Fig.~\ref{fig:alpha_ann_vs_alpha_decay} is easily understood. For higher $ \alpha_{\rm ann} $ values, fewer $ \psi $ particles remain after their freezeout. To compensate, more $ \chi $ particles must be created before $ \psi $'s freezeout, and thus $ \alpha_{\rm decay} $ must be larger. In this sense, this freezeout-and-decay phase is a continuation of the FI phase, with $\chi$ particles generated in different regions of the $\psi$ evolution.

\section{Departure from Kinetic Equilibrium}\label{sec:NKE}

In this section, we discuss non-kinetic equilibrium (NKE) corrections to INDY DM. Thus far, all the results shown and discussed were obtained by solving the integrated BEs \eqref{eq:Bolzmann_equations} for various parameters and initial conditions. Using these equations, we assumed that $ \chi $ is in kinetic equilibrium (KE) with the SM. While this assumption is valid when the decay and inverse decay rates are relatively large (in the coannihilation-via-decays phase), it is not necessarily valid for INDY DM, where the decay coupling is small and the corresponding rates are slow.\footnote{Note that in a given model that realizes the INDY mechanism, additional interactions that can bring the DM into kinetic equilibrium may occur.} (For a discussion of the impact of non-kinetic equilibrium, see {\it e.g} Refs.~\cite{Binder:2017rgn,DAgnolo:2017dbv,Garny:2017rxs}.)

In this section, we relax the assumption of KE after $ x=1 $ and study its impact on INDY DM and the obtained couplings, in the limit of $ Y_{\psi} = Y_{\psi}^{\rm eq} $ ($\alpha_{\rm ann} \rightarrow \infty$). 
We show that the solution of the NKE case provides small corrections to the values of $ \alpha_{\rm decay} $ in the relevant parameter space, rendering the previously presented results valid. We further provide mathematical insight for these results. A reader interested directly in a QFT model realizing the phases discussed thus far can advance directly to the next section.

\subsection{Non-integrated Boltzmann Equation} \label{subsec:FI_general}

We now study the system in the limit of large $\alpha_{\rm ann} $ values, without the assumption of KE. For this, we solve the non-integrated BE for the distribution function $ f_{\chi} $, which is given by 
\begin{multline} \label{eq:unintegrated_BE_for_f_chi}
	\frac{\partial f_{\chi}}{\partial t}-Hp\frac{\partial f_{\chi}}{\partial p} =  \frac{1}{2 E_{\chi}} \intop d\Pi_{\psi} d\Pi_{\phi}  \delta^{(4)}\left(p_{\psi}-p_{\phi}-p_{\chi}\right) \\ 
	 \times \left(2\pi\right)^{4} \overline{\left|\mathcal{M}\right|^{2}}_{\psi\rightarrow\phi\chi} \left(f_{\psi}\left(p_{\psi}\right) - f_{\phi}\left(p_{\phi}\right) f_{\chi}\left(p_{\chi}\right)\right) \, .
\end{multline}

By change of variables from $ \left(t,p_{\chi}\right) $ to $ \left(x,q\right) $ such that $ x = \frac{m_{\chi}}{T\left(t\right)} $ is the same as before and $ q= a(x) p_{\chi} $ is the comoving momentum  (with $ a (x) $ the scale factor), Eq.~\eqref{eq:unintegrated_BE_for_f_chi} turns into a simple ordinary differential equation for each $ q $,
\begin{equation} \label{eq:f_chi_ODE}
	x H \frac{d \bar{f}_{\chi,q}}{dx} = - c\left(x,q\right) \left( \bar{f}_{\chi,q} (x) - \bar{f}_{\chi,q}^{\rm eq} (x) \right)\,,
\end{equation}
where 
\begin{equation} \label{eq:collition_term_NKE}
	c\left(x,q\right) = \frac{2 g_{\psi} m_{\psi}^{2} \alpha_{\rm decay} T }{g_{\chi} p_{\chi} E_{\chi}} e^{-\delta \frac{E_{\chi} }{T}}  \sinh\left( \frac{ \bar{\beta} m_{\psi}^{2} p_{\chi} }{ 2m_{\chi}^{2}T} \right) \, ,
\end{equation}
with the notation $ \bar{f}_{\chi,q} (x) = f_{\chi} \left( t(x), p_{\chi} \left(x,q\right) \right) $, $ \bar{f}_{\chi,q}^{\rm eq} (x) \equiv \exp\left(-\frac{E_{\chi}\left(x,q\right)}{T(x)}\right) $ and
\begin{equation} \label{delta_definition}
	\delta \equiv \frac{m_{\psi}^{2}-m_{\phi}^{2}-m_{\chi}^{2}}{2m_{\chi}^{2}} \, .
\end{equation} 
The complete derivation can be found in Appendix \ref{appx:sec:NKE_BE_derivation} and in Refs.~\cite{DAgnolo:2017dbv,Garny:2017rxs}.

Eq. \eqref{eq:f_chi_ODE} can be integrated in a similar way as the yield in the KE case in Eqs.~\eqref{eq:BE_vertical_branch} and \eqref{eq:vertical_branch_Y_integration}:
\begin{equation} \label{eq:f_chi_integrated_form}
	\bar{f}_{\chi,q} (x) = e^{-\mathbf{A}_{\rm q}(x)} \left( \bar{f}_{q,\chi}\left(1\right) + \intop_{1}^{x}e^{\mathbf{A}_{\rm q}(\eta)}\mathbf{b}_{\rm q}(\eta)d\eta \right) \, 
\end{equation}
with
\begin{equation} \label{eq:a_q_and_b_q_definitions}
	\mathbf{a}_{\rm q}(x)\equiv\frac{c\left(x,q\right)}{x H},\quad
	\mathbf{b}_{\rm q}(x)\equiv\frac{c\left(x,q\right)}{x H}\bar{f}_{\chi,q}^{\rm eq}(x)
\end{equation}
and
\begin{equation} \label{eq:A_q_definition}
	\mathbf{A}_{\rm q}(x)\equiv\intop_{1}^{x}\mathbf{a}_{\rm q}\left(\xi\right)d\xi \, .
\end{equation}
Eqs.~\eqref{eq:a_q_and_b_q_definitions} and ~\eqref{eq:A_q_definition} are analogous to Eqs.~\eqref{vertical_branch_a_b_definitinos} and \eqref{vertical_branch_a_b_definitinos_2}. We assume that at $ x=1 $ all the particles are at kinetic and chemical equilibrium with the SM bath.

Finally, the particle number density $ n_{\chi} (x) $ is given by integration over all momenta, and the yield can be written as
\begin{equation} \label{eq:Y_chi_NKE}
	Y_{\chi}(x) = \frac{1}{s\left(x=1\right)}\frac{g_{\chi}}{2\pi^{2}}\intop q^{2}\bar{f}_{\chi,q}(x)dq \, .
\end{equation}
The integral is dominated by the region of $ q \sim m_{\chi} $. This can be understood by the following arguments: 
\begin{itemize}
    \item For small momenta ($ q \ll m_{\chi} $) the distribution function at $x=1$ is similar ($ \bar{f}_{\chi,q}^{\rm eq} \left(1\right) \sim \bar{f}_{\chi,m_{\chi}}^{\rm eq} \left(1\right)  $) and the collision rate is similar ($ c\left(x,q\right) \sim c\left(x,m_{\chi}\right) $\footnote{Since for small momenta $ E_{\chi} \sim m_{\chi} $ and $ \sinh{\left( \alpha p \right)}/p \approx \alpha $ for $ \alpha p \ll 1 $.}), but the contribution to the integral is small because of the $q^2 dq$
measure.
    \item  For  high momenta ($ q \gg m_{\chi} $) the initial value is exponentially smaller ($ \bar{f}_{\chi,q}^{\rm eq} \left(1\right)\ll \bar{f}_{\chi,m_{\chi}}^{\rm eq} \left(1\right) $), and the collision term exponentially increases with $ q $.
\end{itemize}
Thus, the region $ q \sim m_{\chi} $ is the most dominant one in the integral evaluation.

\begin{figure*}[t]
	\centering	
	\includegraphics[width=\columnwidth]{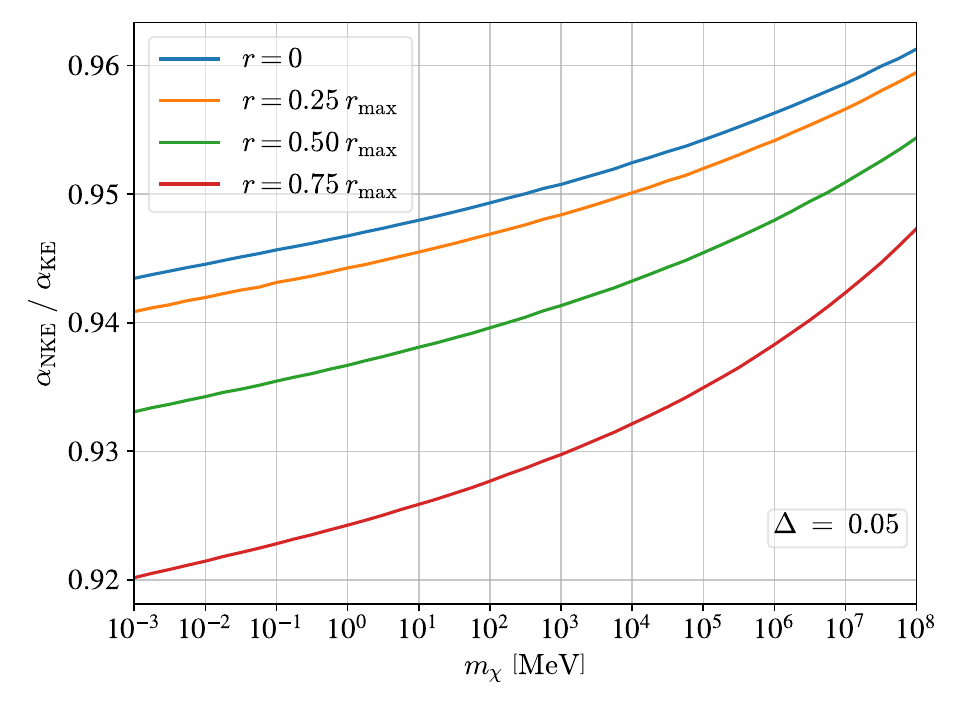}
	\includegraphics[width=\columnwidth]{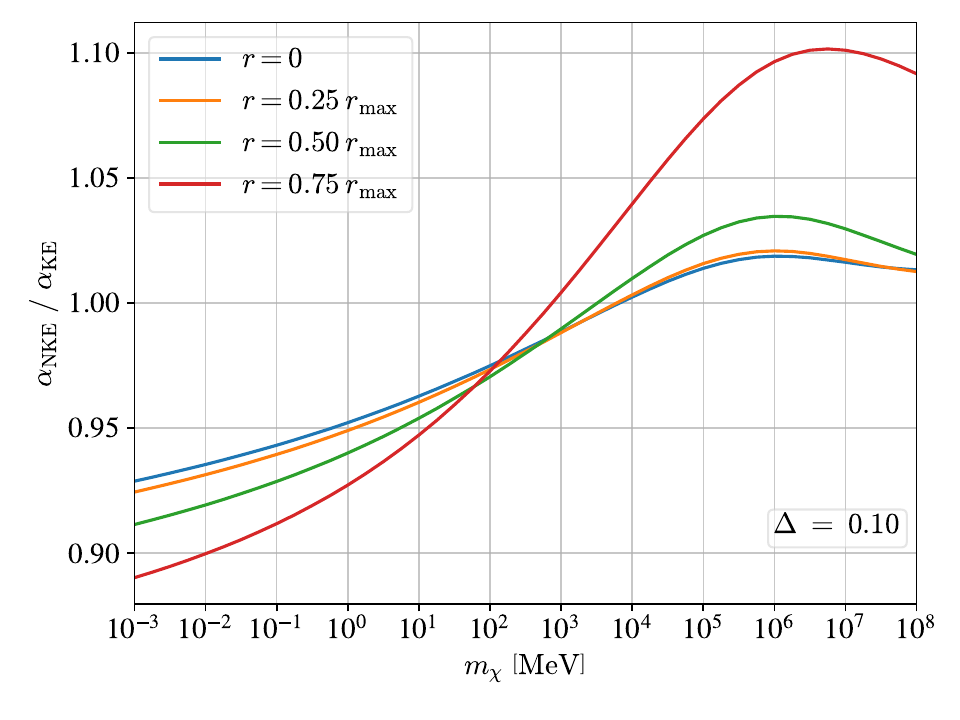}
	\includegraphics[width=\columnwidth]{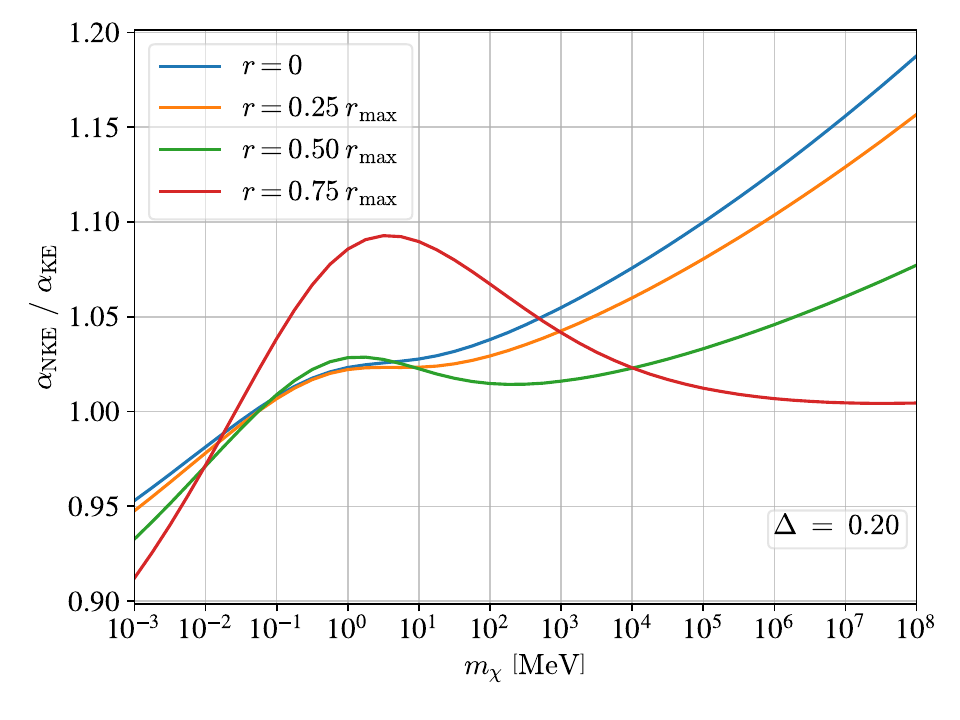}
	\includegraphics[width=\columnwidth]{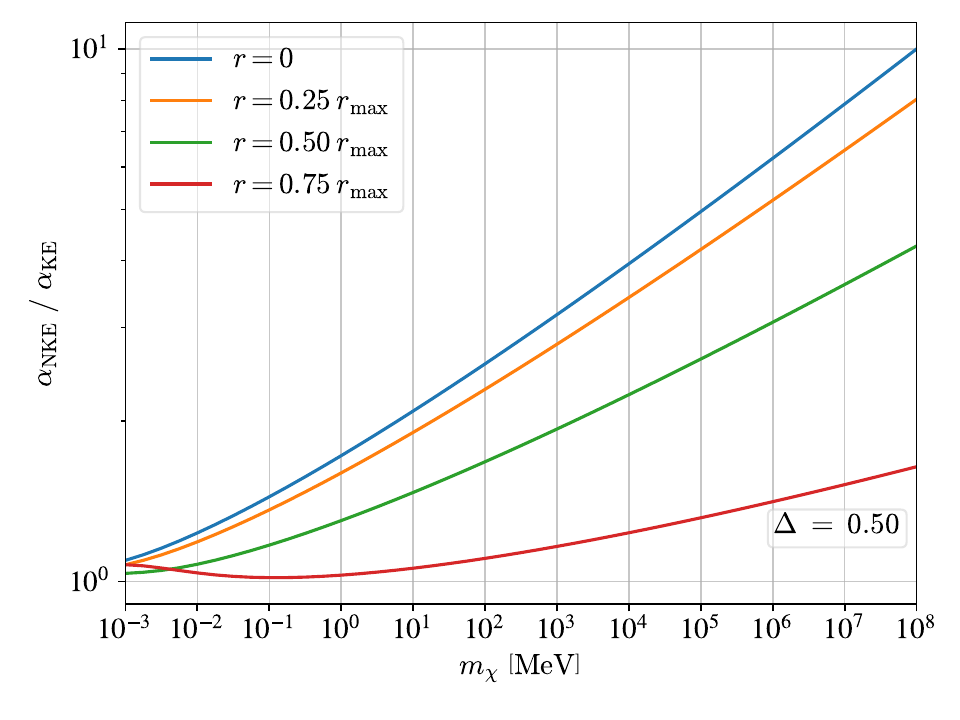}
	\caption{{\bf Impact of kinetic equilibrium.} The ratio between the decay couplings obtained without and with kinetic equilibrium, as a function of  DM mass. The various panels are for different mass splittings $\Delta$, while the curves inside each panel are obtained for different $\phi$ masses, represented by various values of $ r=m_{\phi}/m_\chi$ relative to $r_{\rm max} = \Delta $ (see Eq.~\eqref{eq:delta_and_r_definition}). The calculations in this figure are made with constant $ g_{*} $. Note that for $\Delta=0.5$, unitarity considerations typically limit the DM mass to $m_\chi\lsim 10^4$~MeV, see Fig.~\ref{fig:alpha_ann_vs_mass}. 
    } 
	\label{fig:alpha_in_NKE}
\end{figure*}

\subsection{Numerical Evaluation of the Decay Coupling} \label{subsec:NKE_alpha_evaluation}

Fig.~\ref{fig:alpha_in_NKE} presents the ratio $ \alpha_{\rm NKE}/\alpha_{\rm KE} $, namely the ratio between the couplings that reproduce the observed DM relic abundance without and with kinetic equilibrium, assuming equilibrium initial conditions, as a function of $ m_{\chi}$, for various fixed values of $ \Delta $ and~$ r $.\footnote{Note that the collision function in Eq.~\eqref{eq:collition_term_NKE} has non-trivial dependence on $ m_{\phi} $ and so unlike the KE case, here we explicitly show the results for massive $ \phi $.}

For all the curves with $ \Delta \le 0.2 $, the NKE corrections to the coupling are relatively small (up to $ 20 \% $).
For $ \Delta = 0.5 $, the corrections to the coupling might be up to an order of magnitude for high DM masses $m_\chi\sim 10^7-10^8$~MeV, but such high masses for this large mass splitting are unlikely in the mechanism due to the unitarity bound on the annihilation cross section (see Section~\ref{subsec:horizontal} and Fig.~\ref{fig:alpha_ann_vs_mass}, which restricts to $m_\chi\lsim 10^4$~MeV for large mass splittings). Moreover, the couplings ratio is close to unity around $ 1 \, {\rm keV} $ and grows up to $ 10 $ only after eleven orders of magnitude in DM mass for massless $\phi$ ($r=0$). For massive $\phi$ ($0<r<\Delta$), which is the case in the model we present latter in Section~\ref{sec:qft_model}, the slope in Fig.~\ref{fig:alpha_in_NKE} decreases with the increase of $m_{\phi}$. 
We learn that the NKE corrections to the coupling values are rather small. 

The following paragraphs are dedicated to explaining the proximity of the obtained couplings along with other features in Fig.~\ref{fig:alpha_in_NKE}.
Since Eq. \eqref{eq:f_chi_integrated_form} has the same functional form as Eq.~\eqref{eq:vertical_branch_Y_integration}, comparison between the KE and NKE cases is discussed separately for the two INDY types of DM--- the out of chemical equilibrium case which is relevant for small DM mass and small mass splittings; and the in chemical equilibrium case, relevant for large DM mass or large mass splittings.

\subsubsection{Small Mass Splitting Case}

For low $ \Delta $ values, the initial condition terms in Eqs.~\eqref{eq:vertical_branch_Y_inf} and \eqref{eq:f_chi_integrated_form} dominate the relic abundance. To examine analytically such case and explaining why $ \alpha_{{\rm KE}} \approx \alpha_{{\rm NKE}} $, we look at the ratio between the inverse decay rates $ \mathbf{a}_{\rm q} $ of relevant $ q $ mods ($q \sim m_{\chi}$) to the inverse decay rate of the yield in the KE case $ \mathbf{a} $.

For the KE case, the inverse decay rate in the NR approximation is
\begin{equation} \label{eq:x_H_a_KE}
	x H\mathbf{a}(x) \approx \alpha_{\rm KE} \bar{\beta} m_{\psi} \frac{g_{\psi}}{g_{\chi}} \left(\frac{m_{\psi}}{m_{\chi}}\right)^{3/2} e^{-\Delta x} \, .
\end{equation}
Assuming constant $ g_{*} $ and $ g_{*s} $ for simplicity (which provides $ a = x $ and $ q = p_{\chi} x $), the NKE inverse decay rate is 
\begin{equation} \label{eq:x_H_a_q}
	x H\mathbf{a}_{\rm q}(x) \approx \frac{2g_{\psi}m_{\psi}^{2}\alpha_{\rm NKE}}{g_{\chi} q }  \sinh{\left(\frac{m_{\psi}^{2}q}{2m_{\chi}^{3}}\bar{\beta}\right)}  e^{-\delta x - \frac{\delta}{2x}\frac{q^{2}}{m_{\chi}^{2}}}  \, ,
\end{equation}
where we used NR approximations for the energy -  $ E_{\chi} \approx m_{\chi}+\frac{1}{2}\frac{p_{\chi}^{2}}{m_{\chi}} $ in the exponent, and $ E_{\chi} \approx m_{\chi} $ in the denominator.

Since $ \bar{\beta} \le \bar{\beta}_{m_\phi =0} = \frac{m_{\psi}-m_{\chi}}{m_{\psi}}\frac{m_{\psi}+m_{\chi}}{m_{\psi}} $, the term inside the hyperbolic sine function is small,\footnote{ The term inside the hyperbolic sine in  Eq.~\eqref{eq:x_H_a_q} obeys
	\begin{equation}
		\frac{m_{\psi}^{2}}{2m_{\chi}^{2}} \frac{q}{m_{\chi}} \bar{\beta} \le \frac{q}{m_{\chi}} \frac{m_{\psi}+m_{\chi}}{2m_{\chi}} \Delta \, \sim \Delta \,
	\end{equation}
	because $ \frac{m_{\psi}+m_{\chi}}{2m_{\chi}}\sim 1 $, the concerning momenta are $ q \sim m_{\chi} $, and $ \Delta $ is small.} and a linear approximation can be made, providing 
\begin{equation} \label{eq:x_H_a_q_approx}
	x H\mathbf{a}_{\rm q}(x) \approx \frac{g_{\psi}}{g_{\chi}} \alpha_{\rm NKE} e^{-\delta x - \frac{\delta}{2x}\frac{q^{2}}{m_{\chi}^{2}}} \frac{m_{\psi}^{4}}{m_{\chi}^{3}}\bar{\beta} \, .
\end{equation}

In the case of $ m_{\phi} = 0 $, the ratio between the rates in Eqs. \eqref{eq:x_H_a_q_approx} and \eqref{eq:x_H_a_KE} is
\begin{equation} \label{eq:ratio_a_a_q_massless_phi}
	\frac{\mathbf{a}_{\rm q}(x) }{ \mathbf{a}(x)} \approx \frac{\alpha_{{\rm NKE}}}{\alpha_{{\rm KE}}} e^{-\Delta^{2}x} e^{-\frac{1}{2x}\frac{m_{\psi}+m_{\chi}}{2m_{\chi}}\Delta\frac{q^{2}}{m_{\chi}^{2}}} \left(\frac{m_{\psi}}{m_{\chi}}\right)^{3/2} \, ,
\end{equation}
and substituting $ x_{*} = \Delta^{-1} $ provides
\begin{equation} \label{eq:ratio_a_a_q_massless_phi_x_star}
	\frac{\mathbf{a}_{\rm q}\left(x_{*}\right) }{ \mathbf{a}\left(x_{*}\right)} \approx \frac{\alpha_{{\rm NKE}}}{\alpha_{{\rm KE}}} e^{-\Delta} e^{-\frac{1}{2}\frac{m_{\psi}+m_{\chi}}{2m_{\chi}}\Delta^{2}\frac{q^{2}}{m_{\chi}^{2}}} \left(1+\Delta\right)^{3/2} \, .
\end{equation}
It follows from Eq.~\eqref{eq:ratio_a_a_q_massless_phi_x_star} that the ratio between the rates at $ x \sim x_{*} $ is close to the ratio between the couplings $ \frac{\mathbf{a}_{\rm q}\left(x_{*}\right) }{ \mathbf{a}\left(x_{*}\right)} \sim \frac{\alpha_{{\rm NKE}}}{\alpha_{{\rm KE}}} $. Since this region of $ x $ is contributes the most in evaluating the total probability to inverse decay, we have $ \frac{\mathbf{A}_{\rm q} }{ \mathbf{A}} \sim \frac{\alpha_{{\rm NKE}}}{\alpha_{{\rm KE}}} $. 

Thus, the same probability for a particle to inverse decay is obtained when close values of decay coupling are taken, which provide the same (observed) relic abundance in both KE and NKE cases, as discussed above.

In the massive $ \phi $ case, the ratio is the same as in Eq.~\eqref{eq:ratio_a_a_q_massless_phi}, up to an additional factor of $ \exp \left(\left(\frac{1}{2}x+\frac{1}{4x}\frac{q^{2}}{m_{\chi}^{2}}\right)\frac{m_{\phi}^{2}}{m_{\chi}^{2}}\right)$.
For $ x= x_{*} $, this factor equals
\begin{equation} \label{eq:x_H_a_q_massive_phi_factor}
	e^{\left(\frac{1}{2}x_{*} + \frac{1}{4x_{*}}\frac{q^{2}}{m_{\chi}^{2}}\right) \frac{m_{\phi}^{2}}{m_{\chi}^{2}}} = e^{\frac{1}{2\Delta}\frac{m_{\phi}^{2}}{m_{\chi}^{2}}} e^{\frac{\Delta}{4}\frac{q^{2}}{m_{\chi}^{2}}\frac{m_{\phi}^{2}}{m_{\chi}^{2}}} \, .
\end{equation}
Since $ \frac{1}{\Delta}\frac{m_{\phi}}{m_{\chi}} < 1 $, we obtain $ \frac{1}{\Delta}\left(\frac{m_{\phi}}{m_{\chi}}\right)^{2}<\Delta $ and $ \frac{\Delta}{4}\frac{q^{2}}{m_{\chi}^{2}}\frac{m_{\phi}^{2}}{m_{\chi}^{2}}<\frac{\Delta^{3}}{4}\frac{q^{2}}{m_{\chi}^{2}} $ and so the two exponents are small. Again, the ratio between the rates is close to the ratio between the couplings, and the same augment as in the massless $ \phi $ case is valid, providing $ \alpha_{\rm NKE} \sim \alpha_{\rm KE} $.  Moreover, since in Eq.~\eqref{eq:x_H_a_q_massive_phi_factor} the additional factor increases with $ m_{\phi} $, $ \alpha_{\rm NKE} $ should decrease to obtain the same rate, thus explaining why in Fig.~\ref{fig:alpha_in_NKE} the $ \Delta = 0.05 $ curves decrease slightly as $ r $ increases.

As a final remark, we note that the maximum of $ \mathbf{a}_{\rm q} $ is $ q $ dependent and not exactly at $ x_{*} $ , but the qualitative analysis presented above is not affected by the slight maximal rate time change.

\subsubsection{Large Mass Splitting Case} \label{subsubsec:NKE_in_ce}
Here, we aim to understand why $ \alpha_{{\rm NKE}} \sim \alpha_{{\rm KE}} $ even for relatively high $ \Delta $ values. In this case, $ \psi $'s decays dominate the relic abundance. Using the saddle point approximation, it can be shown that the phase space distribution function dependence on the mass scale (up to logarithmic corrections) is 
\begin{equation} \label{eq:f_chi_in_CE_approx}
	\bar{f}_{\chi,q,\infty} \sim \left( \frac{ \alpha_{\rm NKE}m_{\rm pl} }{ q } \right)^{-\frac{1}{\delta}} \, .
\end{equation}
The complete derivation is given in Appendix \ref{appx:sec:saddle_point}. Using Eq.~ \eqref{eq:Y_chi_NKE}, the relic abundance's dependence on $ m_\chi $ is
\begin{equation} \label{eq:Y_chi_NKE_in_CE_approx}
	Y_{\chi,\infty} \sim \left (\frac{\alpha_{\rm NKE}m_{\rm pl}}{m_{\chi}} \right)^{-\frac{1}{\delta}} \, ,
\end{equation}
and by matching the relic abundance in Eq.~\eqref{eq:Y_chi_NKE_in_CE_approx} with the observed value we produce the power-law relation 
\begin{equation} \label{eq:alpha_NKE_as_x}
	\alpha_{\rm NKE} \sim m_{\chi}^{1+\delta}.
\end{equation}

Finally, Comparing the NKE and KE cases (Eqs.~ \eqref{eq:alpha_in_CE_INDY} and \eqref{eq:alpha_NKE_as_x}) provides
\begin{equation} \label{eq:alpha_NKE_to_NE_in_CE_power_law}
	\frac{\alpha_{\rm NKE}}{\alpha_{\rm KE}} \sim m_{\chi}^{\delta-\Delta} = m_{\chi}^{\frac{\Delta^{2}}{2} - \frac{r^{2}}{2}} \, .
\end{equation}
This remarkable result explains the low growth rate of the ratio for high $ \Delta $ values observed in Fig.~\ref{fig:alpha_in_NKE}. For $ \Delta = 0.5, r=0 $, the expected power is $ \frac{1}{8} $, which is a small power by itself, and close to the slope in the figure, $ \sim \frac{1}{11} $. For $ \Delta = 0.2 $, the expected power reduces to an even smaller number $ \frac{1}{50} $, explaining the growth of only $\sim25\% $ after eleven orders of magnitude. 

Additionally, increasing $ r $ reduces the power, in correspondence with the $ \Delta = 0.5 $ curves at Fig.~\ref{fig:alpha_in_NKE}. For $ \Delta = 0.1,0.2 $, the power is so low that the leading term in Eq.~\eqref{eq:Y_chi_NKE_in_CE_approx} is not enough to explain the results, and a more subtle approach is needed.

In summary, we have verified that the results for $ \alpha_{\rm decay} $ 
for INDY DM, which were obtained in the previous sections assuming kinetic equilibrium, are in close proximity to the obtained values in the absence of such an assumption, rendering our analysis valid.

\section{Model}\label{sec:qft_model}

The system considered in this work, described by the BEs~\eqref{eq:Bolzmann_equations}, can be realized by a simple renormalizable toy model~\cite{Frumkin:2021zng}. We now present such a model and discuss its relevant parameter space which manifests the phases of INDY DM and coannihilation via decays, and discuss possible phenomenology. 

We consider a dark sector with internal gauge symmetry $U(1)_{d}$ with gauge field $ A_{d} $ (dark photon) and gauge coupling $ e_{d} $ (dark charge). 
It contains two Dirac fermions $ \chi $ and $ \psi $, and a complex scalar $ H_{d} $ with charges $0$, $1$, and $1$ respectively.
Note that $ \chi $ and $ \psi $ take the same role as in the previous sections ($ \psi $ decays to $ \chi $), while $ H_{d} $ does not explicitly appear in the mechanism as presented thus far.
The general renormalizable Lagrangian contains mass terms for the fermions 
\begin{equation}
    \mathcal{L}_{\rm Dirac}= -m_\chi \bar{\chi} \chi-m_\psi \bar{\psi} \psi\,,
\end{equation} 
spontaneous symmetry breaking (SSB) potential for the scalar
\begin{equation}
	V\left(\phi\right)= -\frac{1}{2}\mu^{2}H_{d}^{*}H_{d} +\frac{1}{4}\lambda\left(H_{d}^{*}H_{d}\right)^{2} \, ,
\end{equation}
and a Yukawa interaction
\begin{equation}
	\mathcal{L}_{\text{Yukawa}} = - yH_{d}^{*}\bar{\chi}\psi+h.c \, .
\end{equation}
The connection to the SM is through mixing between the dark photon and the photon ($ A $) fields,
\begin{equation}
	\mathcal{L} \supset - \frac{\epsilon}{2} F_{d}^{\mu\nu}F_{\mu\nu} \, .
\end{equation}

The field $ H_{d} $ acquires a vacuum expectation value $ v_{H_{d}} \equiv \left\langle H_{d} \right\rangle = \frac{\mu}{\sqrt{\lambda}} $ through SSB, providing a real scalar field $ h_{d}$ with $ \left\langle h_{d} \right\rangle = 0 $ and $ m_{h_{d}} = \mu $, and mass term to the dark photon of \cite{Peskin:1995ev,Cheng:1984vwu} 
\begin{equation} \label{eq:m_d_from_SSB}
	m_{A_d}^{2}=2e_{d}^{2}v_{H_{d}}^{2} \, .
\end{equation}
Additionally, the Yukawa interaction term after SSB yields a mass between the fermions $ \mathcal{L} \supset - yv_{H_{d}}\bar{\chi}\psi+h.c$. A rotation to the mass diagonalized fields $ \chi'$ and $\psi' $, using the rotation angle $ \vartheta $ defined by
\begin{equation}
	\sin 2 \vartheta = \frac{ 2yv_{H_{d}} }{ \left(m_{\psi'}-m_{\chi'}\right) } \, ,
\end{equation}
removes the mass mixing. The term $ \mathcal{L} \supset -e_{d}\bar{\psi}\slashed{A}_{d}\psi $ after the rotation yields an explicit interaction between $ \psi' $, $ \chi' $ and $ A_{d} $: 
\begin{equation}
	\mathcal{L} \supset  e_{d}\cos\vartheta\sin\vartheta\bar{\chi}'\slashed{A}_{d}\psi' + h.c \, .
\end{equation}

Since only the mass eigenstates ($ \chi'$ and $\psi' $) are considered, in what follows we neglect the prime in denoting the fields.

\subsection{Parameter restrictions} \label{subsec:model_restrictions}

We impose several conditions on the parameters of the model, such that the  decay $ \psi \rightarrow \chi A_{d} $ and annihilation $ 2\psi \rightarrow 2A_{d} $ processes dominate the relic abundance.

First, for the decay to the dark photon to occur, the restriction $ m_{\chi} + m_{A_d} < m_{\psi} $ must be imposed. 
Additionally, to prevent the decay to the dark Higgs $\psi \rightarrow \chi h_{d} $ from occurring (else it would dominate the relic abundance), we take $ m_{\chi} + m_{h_d} > m_{\psi} $. The inequalities over the masses can be rewritten as a constraint over the dark charge
\begin{equation} \label{eq:e_d_constraint_1}
	e_{d} < \sqrt{\frac{\lambda}{2}} \min \left\{  \frac{m_{\psi}-m_{\chi}}{m_{h_d}} \; , \frac{m_{A_{d}}}{m_{\psi}-m_{\chi}} \right\} \,.
\end{equation}
Next, for the dark sector to thermalize with the SM, the reaction rate $ A_d \leftrightarrow e^{+}e^{-} $ must be fast enough along the evolution of the DM abundance. This provides a lower bound on $ \epsilon $ which is imposed later in this section by requiring $ \Gamma_{A_{d} \rightarrow e^{+}e^{-}} \geq H $ at $ x=1 $.

Finally, $ h_{d} $ must follow equilibrium (to avoid co-scattering of $\chi $ with $ h_{d} $ that is out of CE) and decay into lighter particles (otherwise it is a DM candidate by itself). This can be achieved in two ways. If $ e_d < \frac{1}{2} \sqrt{\frac{\lambda}{2}} $ the reaction $ h_{d} \rightarrow 2 A_{d} $ is permitted and no further restriction is added. If $ \frac{1}{2} \sqrt{\frac{\lambda}{2}} < e_d < \sqrt{\frac{\lambda}{2}} $, the reaction $ h_{d} \rightarrow A_{d} e^{+}e^{-} $ is responsible for the dark Higgs decay. For $h_{d}$ to follow CE in the latter case, we demand $ \Gamma_{h_{d} \rightarrow A_{d} e^{+}e^{-}} \geq H $ at $ T=m_{h_{d}} $, which provides again a lower bound on $ \epsilon $.

\begin{figure*}[ht!]
	\centering
	\includegraphics[width=\columnwidth]{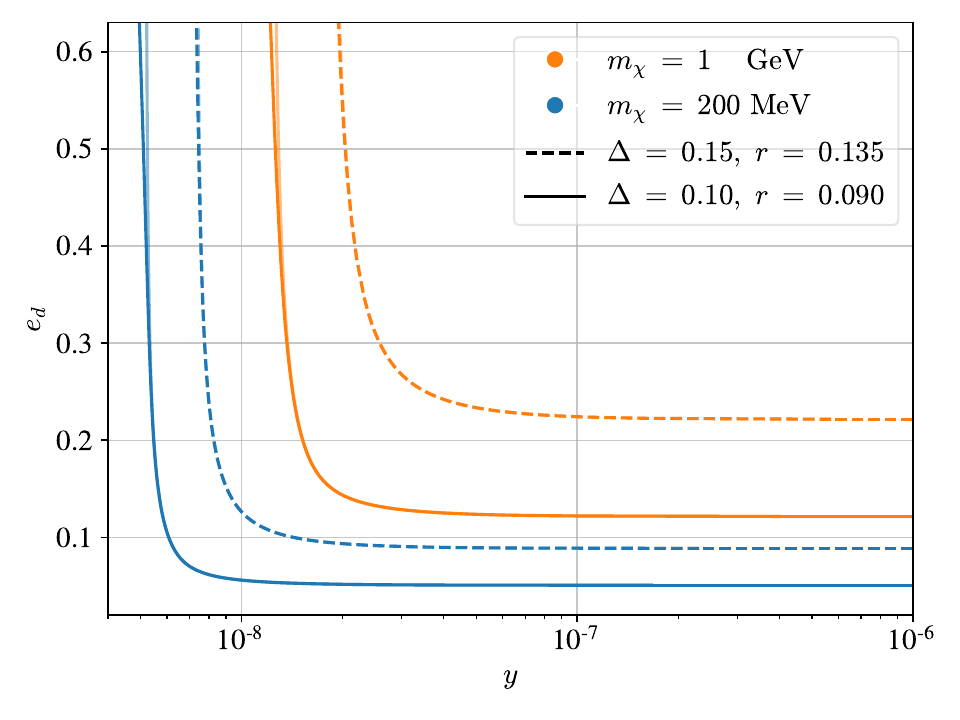}
	\includegraphics[width=\columnwidth]{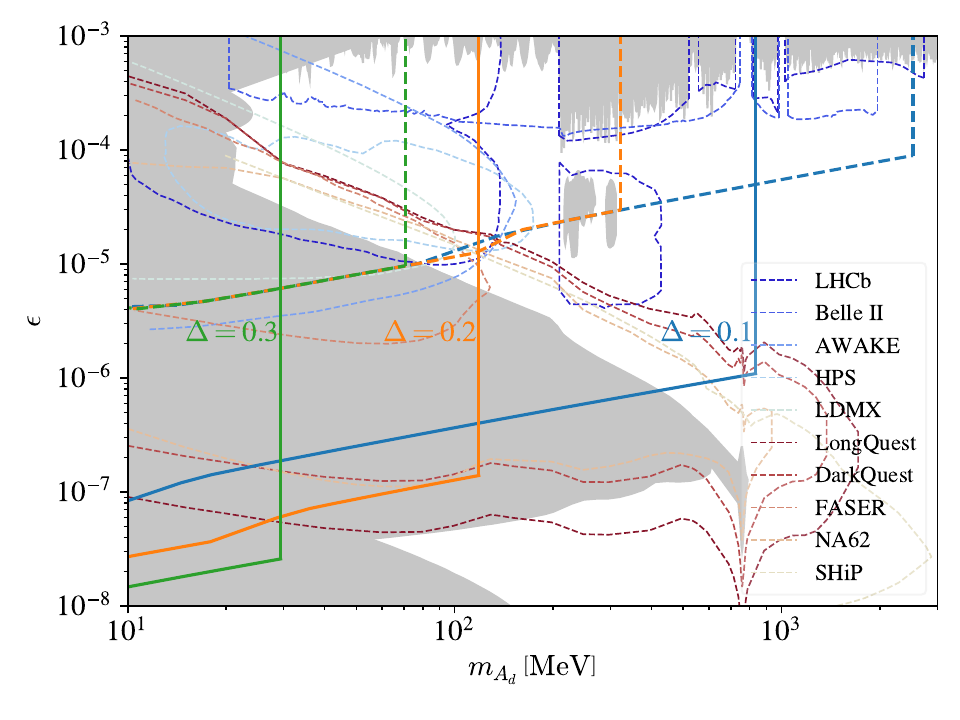}
	\caption{{\bf Model.} {\it Left:} The dark charge $ e_{d} $ verses the Yukawa coupling $ y $, which reproduces the observed abundance for the model, for different DM masses. The lighter colored solid curves in the plot show the values obtained with decay and annihilation reactions only, without co-scattering. Here we set $ \lambda = 1 $. {\it Right:} Allowed dark photon parameter space. Solid thick colored curves indicate the minimal kinetic mixing as a function of dark photon mass within the model, for various values of mass splitting $\Delta$ with $ \lambda = 1, \mu = 2 \left( m_{\psi} - m_{\chi} \right) , e_d = 0.32 $. In this case, the constraint on $ \epsilon $ comes from the thermalization of the sectors through the $ A_d \leftrightarrow e^{+}e^{-} $ reaction. The dashed thick lines indicate the minimal kinetic mixing for different parameter choice: $ \lambda = 1, \mu = 1.1 \left( m_{\psi} - m_{\chi} \right) , e_d = 0.58 $. Here, the constraint on $ \epsilon $ comes from the $ h_{d} \rightarrow A_{d} e^{+}e^{-} $ reaction rate (see Section~\ref{subsec:model_restrictions}). In both cases the connection between the masses of the dark photon and the DM is $m_{A_d} = 0.9\, \Delta m_\chi$. We show existing limits~\cite{Fradette:2014sza,Alexander:2016aln,Chang:2016ntp,Hardy:2016kme,Pospelov:2017kep,NA64:2018lsq,LHCb:2017trq,LHCb:2019vmc,Parker:2018vye,Tsai:2019buq} in shaded gray, with future projections~\cite{Celentano:2014wya,Belle-II:2018jsg,Ilten:2015hya,Alekhin:2015byh,Ilten:2016tkc,Alexander:2016aln,Caldwell:2018atq,Berlin:2018pwi,Berlin:2018bsc,FASER:2018eoc,NA62:2312430,Tsai:2019buq} indicated by the colored thin dashed curves. 
    }
	\label{fig:qft_model_parameter_space}
\end{figure*}

\subsection{Interactions} \label{subsec:model_reactions}

We now relate the model's parameters to the couplings $ \alpha_{\rm decay} $ and $ \alpha_{\rm ann} $ in Eq.~\eqref{eq:gamma_parameterization} and Eq.~\eqref{eq:sigma_v_parameterization}. 

At the tree-level approximation, the decay coupling is given by \cite{Shtabovenko:2020gxv,Shtabovenko:2016sxi,Mertig:1990an}
\begin{equation} \label{eq:alpha_decay_as_y}
	\alpha_{\rm decay} = \frac{y^{2}\left(\left(m_{\chi}-m_{\psi}\right)^{2}-m_{A_d}^{2}\right)\left(\left(m_{\chi}+m_{\psi}\right)^{2}+2m_{A_d}^{2}\right)}{32\pi m_{\psi}^{2}\left(m_{\chi}-m_{\psi}\right)^{2}}\, .
\end{equation}
Since the relevant values of $ \alpha_{\rm decay} $ are very small (see Fig.~\ref{fig:alpha_decay_as_function_of_m_chi_and_w}), $ y $ must be small as well. 

By neglecting diagrams and terms with high orders of $ y $, calculating the thermally averaged cross-section in the s-wave approximation provides \cite{Shtabovenko:2020gxv,Shtabovenko:2016sxi,Mertig:1990an}
\begin{equation} \label{eq:alpha_ann_as_e_d}
	\alpha_{\rm ann}^{2}=\frac{e_{d}^{4}\sqrt{1-\frac{m_{A_d}^{2}}{m_{\psi}^{2}}}\left(m_{\psi}^{2}-m_{A_d}^{2}\right)}{32\pi\left(m_{\psi}^{2}-\frac{1}{2}m_{A_d}^{2}\right)^{2}}m_{\psi}^{2} \, .
\end{equation}

Additional reactions involving $ \chi $ occur in the model, but their effect is small. Since $ y $ is small, all the reactions with matrix element squared of order $ \mathcal{O} \left( y^3 \right) $ or above are insignificant, leaving only nine reaction types with $\overline{\left|\mathcal{M}\right|^2} \propto y^2 $. In the following, we show that the affect of those nine reactions is small as well. 

The fastest of the nine interactions is co-scattering with the dark photon $ \chi A_{d} \rightarrow \psi A_{d} $, since it is proportional to the dark photon number density $ n_{A_d} $. The ratio between its rate to inverse decay is 
\begin{equation} \label{eq:model_rate_ratio_INDY_coscattering}
	\frac{\text{Rate}_{\chi A_d \rightarrow \psi A_d}}{\text{Rate}_{\chi A_d \rightarrow \psi }} = \frac{\left\langle \sigma v\right\rangle _{\psi A_{d}\rightarrow\chi A_{d}} n_{A_{d},{\rm eq}}}{\left\langle \Gamma\right\rangle _{\psi\rightarrow\chi A_{d}}} \sim \frac{n_{A_{d},{\rm eq}}}{m_{\psi}^{3}} \, .
\end{equation}
In the parameter space considered in this work, the dark photon mass is settled near its threshold $ m_d \lesssim m_{\psi} - m_{\chi} $ (see next subsection). From Eq.~ \eqref{eq:model_rate_ratio_INDY_coscattering}, the ratio between the rates (and the co-scattering rate itself) decreases rapidly at $ x \sim \frac{m_{\chi}}{m_{A_d}} $, which is near the temperature at which the inverse decay rate is maximal $ x_{*} = \Delta^{-1} $ (see Eq.~\eqref{eq:a_of_x_NR}). Thus, the inverse decays dominate the freeze-out process in such parameter choices, and co-scattering serves as a secondary process. 

It is worth mentioning that since for $ x<x_{*} $ the co-scattering rate might be higher, it can change the `effective' initial conditions for the inverse decay process (a possibility mentioned in \cite{Garny:2017rxs}); however, we have not found significant change to the obtained coupling $ y $ in the calculations done for this work. The small effect of initial conditions on $y$ is in correspondence with the results shown in Section~\ref{subsec:vertical}, which discusses the effect on changing the initial conditions on $\alpha_{\rm decay}$.

The remaining eight reactions have a lower rate compared to co-scattering with the dark photon. The rates of co-scattering involving the scalar $ h_{d} $ ($ \chi A_{d}\rightarrow\psi h_{d} $, $ \chi h_{d}\rightarrow\psi h_{d}  $ and $ \chi h_{d}\rightarrow\psi A_{d} $) are slower than the rate of co-scattering with $ A_{d} $ only by a factor of~$\sim n_{ h_{d}}^{\rm eq}/ n_{A_{d}}^{\rm eq} $ or~$\sim\left(n_{ h_{d}}^{\rm eq}/ n_{A_{d}}^{\rm eq} \right)^{2}$. The reactions with rate containing a factor of $ n_{\psi} $ ($ \chi\psi \rightarrow \psi\psi $ , $ \chi\bar{\psi} \rightarrow \psi\bar{\psi} $, $ \chi\bar{\psi} \rightarrow A_{d}A_{d} $, $ \chi\bar{\psi} \rightarrow  h_{d} A_{d} $ and $ \chi \bar{\psi} \rightarrow  h_{d} $) are exponentially suppressed as well ($ \chi \bar{\psi} \rightarrow  h_{d} $ is also forbidden in our parameter space choice).

\subsection{Results} \label{subsec:model_benchmark}

For the model to avoid cosmological constraints from $ N_{\rm eff} $ and Big Bang Nucleosynthesis \cite{Parker:2018vye,Banerjee:2019pds,BaBar:2017tiz,Andreas:2012mt,BLUMLEIN2014320,Batley_2015178,Tsai:2020vpi,Ibe_2020,Sabti_2020,Bauer:2018onh}, the parameters are chosen such that the dark photon (which is chosen to be the lightest particle) has mass of $ m_{A_d} \ge 18 \ {\rm MeV} $. High values of $ \alpha_{\rm ann} $ are needed for high mass scales (see Fig.~\ref{fig:alpha_ann_vs_mass}), which translates to a requirement of high  $ e_{d} $ values and large $ m_{A_{d}} $ masses (see Eqs.~\eqref{eq:alpha_ann_as_e_d} and~\eqref{eq:m_d_from_SSB}).

The left panel of Fig.~\ref{fig:qft_model_parameter_space} shows $ y $ and $ e_d $ values that reproduce the observed DM relic abundance for different DM masses, mass splittings and values of $r$. Co-scattering with the dark photon plays little role, as is evident from the proximity of the results with and without co-scattering.
The analogy of Fig.~\ref{fig:qft_model_parameter_space} to Fig.~\ref{fig:alpha_ann_vs_alpha_decay}  emphasizes how the model manifests the INDY and coannihilation via decay mechanisms, through Eqs.~\eqref{eq:alpha_ann_as_e_d} and~\eqref{eq:alpha_decay_as_y}, for various parameters.

INDY DM can be expected to evade  conventional searches such as direct and indirect detection of DM. The reason is that in the dark sector, all $ \chi $ interactions are suppressed by factors of $y\sim 10^{-8}$, and for interaction with the SM, additional factors of $\epsilon \ll 1$ are present. 
For example, the obtained cross-section of $ \chi e \rightarrow \chi e $ elastic scattering, which is given by \cite{Shtabovenko:2020gxv,Shtabovenko:2016sxi,Mertig:1990an,Peskin:1995ev}
\begin{equation} \label{eq:cross_section_chi_e}
	\sigma_e=\frac{y^{4}\epsilon^{2}e^{2}m_{e}^{2}m_{\chi}^{2}}{4\pi e_{d}^{2}\left(m_{e}+m_{\chi}\right)^{2}\left(m_{\chi}-m_{\psi}\right)^{4}}\, ,
\end{equation}
where $ m_{e} $ and $ e $ are the mass and electric charge of the electron, is 
smaller by several orders of magnitude than current reach of direct direction experiments \cite{Kuflik:2017iqs,Alexander:2016aln,Essig:2017kqs,Liu:2017drf,Essig:2022dfa}. 
Note that while Eq.~\eqref{eq:cross_section_chi_e} is model dependent, the low values of $ \alpha_{\rm decay} $ obtained in the general case (Fig.~\ref{fig:alpha_decay_as_function_of_m_chi_and_w}) suggests that $ \chi $ interactions are expected to be weak in any model which realizes the mechanism.

The dark photon can, however, be searched for directly. 
The right panel of Fig.~\ref{fig:qft_model_parameter_space} shows the minimal value of $ \epsilon $ needed for each $ m_{A_{d}} $, for three choices of parameters within the model. 
Experimental constraints on a visibly decaying dark photon in the mass range $\mathcal{O}(10-1000)$~MeV currently constrain $\epsilon\lesssim {10^{-3}-10^{-4} }$~\cite{Fradette:2014sza,Alexander:2016aln,Chang:2016ntp,Hardy:2016kme,Pospelov:2017kep,NA64:2018lsq,LHCb:2017trq,LHCb:2019vmc,Parker:2018vye,Tsai:2019buq}, and most of the relevant parameter space is expected to be probed by future experiments, as demonstrated in the plot ~\cite{Celentano:2014wya,Belle-II:2018jsg,Ilten:2015hya,Alekhin:2015byh,Ilten:2016tkc,Alexander:2016aln,Caldwell:2018atq,Berlin:2018pwi,Berlin:2018bsc,FASER:2018eoc,NA62:2312430,Tsai:2019buq} (see Ref.~\cite{Tsai:2020vpi} for a recent compilation of existing searches and future projections for visibly decaying dark photons).
Additionally, the INDY dark matter candidate should be detectable through  the direct production of $\psi$ particles that decay into $\chi$ and either additional  invisible products or SM particles. We leave a detailed study of such a signal to future work. 

\section{Conclusions}\label{sec:conclusions}

In this work, we studied in detail the possibility that inverse decays play a meaningful role in setting the abundance of DM. We considered a framework of inverse decay interactions along with a self-annihilation process of the unstable particle, and examined the different phases which emerge in this setup. This includes a coannihilations via decay phase, in which the freezeout of the annihilation determines the relic abundance; an INDY phase, in which the freezeout of inverse decays determines the relic abundance; a FI phase, in which only the unstable particle decays account for the observed DM abundance; a FIFO phase, in which the DM abundance is given by freeze in and then freezeout of the DM; and a freezeout and decay phase, in which the decay plays a role only after the unstable particle freezes out.

We showed both numerically and by analytical approximations that when inverse decay reactions control the relic abundance of DM, much smaller couplings are required compared to other mechanisms such as the WIMP. (Interestingly, a chain of inverse decays can allow very heavy DM with larger couplings of ${\cal O}(1)$; for details see Ref.~\cite{Frumkin:2022ror}.) 
We further verified that relaxing the  assumption of kinetic equilibrium does not impact (conceptually or numerically) the couplings or parameter space of INDY~DM. 
Finally, we presented a renormalizable model that can realize the phases of DM described in this paper 
We mapped out the relevant parameter space within the model for the coannihilation via decays and INDY mechanisms presented in this work, and discussed the possible phenomenology for such DM. There is much room for further model-building efforts where inverse decays control the DM abundance. Interesting experimental signals at accelerators are anticipated from the long-lived particles of such theories, which could be used as a probe for inverse decaying dark sectors. 
 
\begin{acknowledgments}
  The work of YH is supported by the Israel Science Foundation (grant No. 1818/22), by the Binational Science Foundation (grants No. 2018140 and No. 2022287) and by an ERC STG grant (``Light-Dark'', grant No. 101040019). The work  of EK was supported by the Israel Science Foundation (grant No. 1111/17) and by the Binational Science Foundation  (grants No. 2016153 and 2020220). 
The work of H.\,M.\ is supported by the Director, Office of Science, Office of High Energy Physics of the U.S. Department of Energy under the Contract No. DE-AC02-05CH11231, by the NSF grant PHY-2210390, by the JSPS Grant-in-Aid for Scientific Research JP23K03382, MEXT Grant-in-Aid for Transformative Research Areas (A) JP20H05850, JP20A203, Hamamatsu Photonics, K.K, and Tokyo Dome Corportation. In addition, H.\,M.\ is supported by the World Premier International Research Center Initiative (WPI) MEXT, Japan.
This project has received funding from the European Research Council (ERC) under the European Union’s Horizon Europe research and innovation programme (grant agreement No. 101040019).  Views and opinions expressed are however those of the author(s) only
and do not necessarily reflect those of the European Union. The European Union cannot be held responsible for them. 
\end{acknowledgments}

\appendix

\section{Saddle Point Approximation for INDY DM In Chemical Equilibrium} \label{appx:sec:saddle_point}

In this Appendix, we show how we use the saddle point approximation to obtain Eqs.~\eqref{eq:Y_inf_indy_in_CE} and~\eqref{eq:f_chi_in_CE_approx}.

The relic abundance with kinetic equilibrium for an INDY DM that is in CE is dominated by the $ \psi $'s decays term of Eq.~\eqref{eq:vertical_branch_Y_inf},
\begin{equation} \label{eq:appx:y_end_in_CE_INDY}
	Y_{\chi,\infty} = \intop_{1}^{\infty}e^{-\intop_{\eta}^{\infty}\mathbf{a}\left(\xi\right)d\xi}\mathbf{b}(\eta)d\eta\, ,
\end{equation}
where the NR approximations for $ \mathbf{a}(x) $ and $ \mathbf{b}(x) $ can be written as
\begin{equation} \label{eq:appx:a_0_b_0_definition}
	\mathbf{a}(x) = \mathbf{a}_{0}xe^{-\Delta x} , \quad \mathbf{b}(x) = \mathbf{b}_{0}x^{5/2}e^{-\left(1+\Delta \right)x}
	 \, 
\end{equation}
with 
\begin{eqnarray} \label{eq:appx:a_0_and_b_0_value}
	\mathbf{a}_{0} & = & \frac{3}{\pi}\sqrt{\frac{10}{g_{*}}}\frac{g_{\psi}}{g_{\chi}}\bar{\beta}\left(1+\Delta \right)^{5/2}\frac{\alpha_{\rm decay}m_{\rm pl}}{m_{\chi}} \, , \\
	\mathbf{b}_{0} & = & \frac{3^{3}\cdot5^{3/2}}{2^{2}\pi^{9/2}}\frac{g_{\psi}}{g_{*s}\sqrt{g_{*}}}\bar{\beta}\left(1+\Delta \right)^{5/2}\frac{\alpha_{\rm decay}m_{\rm pl}}{m_{\chi}} \, , \nonumber
\end{eqnarray} 

Substituting the above into Eq.~\eqref{eq:appx:y_end_in_CE_INDY} yields
\begin{equation} \label{eq:appx:y_end_in_CE_INDY_2}
	Y_{\chi,\infty} = \mathbf{b}_{0}\intop_{1}^{\infty}e^{-\frac{\mathbf{a}_{0}}{\Delta^{2}}e^{-\eta\Delta }\left(1+\eta\Delta \right)-\left(1+\Delta \right)\eta+\frac{5}{2}\ln\eta}d\eta \, .
\end{equation}
This integral can be approximated by the saddle point approximation,
\begin{equation} \label{eq:appx:saddle_point}
	\intop_{-\infty}^{\infty}e^{f(x)}d\eta 
	\approx 
	e^{f\left(\eta_{s}\right)}\sqrt{\frac{2\pi}{-f''\left(\eta_{s}\right)}} \, ,
\end{equation}
where we note $ \eta_{s} $ as the distinct maxima of $ f (x) $: $ \frac{d}{d\eta}f\left(\eta_{s}\right) = 0 $.  As long as $ \eta_{s} >1 $, which is the case for an INDY that is in CE, the lower bound of the integral in Eq.~\eqref{eq:appx:y_end_in_CE_INDY_2} can be extended from $ 1 $ to $ -\infty $ and Eq.~\eqref{eq:appx:y_end_in_CE_INDY_2} has the same form as Eq.~\eqref{eq:appx:saddle_point} with
\begin{equation} \label{eq:appx:f_saddle_KE_CE}
	f(\eta)=-\frac{\mathbf{a}_{0}}{\Delta^{2}}e^{-\eta\Delta }\left(1+\eta\Delta \right)-\left(1+\Delta \right)\eta+\frac{5}{2}\ln\eta \, .
\end{equation}
The saddle point $ \eta_{s} $ obeys
\begin{equation} \label{eq:appx:saddle_point_value}
	e^{-\left(1+\Delta\right)\eta_{s}} = \mathbf{a}_{0}^{-\frac{\left(1+\Delta\right)}{\Delta}} \left(\frac{\left(1+\Delta\right)-\frac{5}{2}\frac{1}{\eta_{s}}}{\eta_{s}}\right)^{ \frac{\left(1+\Delta\right)}{\Delta}} \, ,
\end{equation}
and using Eqs.~\eqref{eq:appx:y_end_in_CE_INDY_2}, \eqref{eq:appx:saddle_point}, \eqref{eq:appx:f_saddle_KE_CE} and \eqref{eq:appx:saddle_point_value} one can approximate the relic abundance to
\begin{multline}
	Y_{\chi,\infty} \approx \mathbf{b}_{0}\mathbf{a}_{0}^{-\frac{\left(1+\Delta\right)}{\Delta}} \eta_{s}^{\frac{5}{2}-\frac{\left(1+\Delta\right)}{\Delta}} \left(\left(1+\Delta\right)-\frac{5}{2}\frac{1}{\eta_{s}}\right)^{\frac{\left(1+\Delta\right)}{\Delta}}\times \\
	e^{-\frac{1}{\Delta^{2}\eta_{s}}+\frac{5}{2}\frac{1}{\Delta^{2}\eta_{s}^{2}}-\frac{1}{\Delta}-1+\frac{3}{2}\frac{1}{\Delta\eta_{s}}}\sqrt{\frac{2\pi}{\Delta+\Delta^{2}+\frac{5}{\eta_{s}^{2}}-\frac{7}{2}\frac{\Delta}{\eta_{s}}-\frac{1}{\eta_{s}}}} \, .
\end{multline}
Since the dependence of $ \eta_{s} $  on $ m_{\chi} $ is logarithmic (because $ m_{\chi} $ appears only in $ \mathbf{a}_{0} $ in Eq.~\eqref{eq:appx:saddle_point_value}), the dependence of $ Y_{\chi,\infty} $ on the DM mass (up to logarithmic corrections) is given by
\begin{equation}
	Y_{\chi,\infty} \sim \mathbf{b}_{0}\mathbf{a}_{0}^{-\frac{\left(1+\Delta\right)}{\Delta}} \sim \left(\frac{\alpha_{\rm decay}m_{\rm pl}}{m_{\chi}}\right)^{-\frac{1}{\Delta}} ,
\end{equation}
which is Eq.~\eqref{eq:Y_inf_indy_in_CE} of  the main text.

The case without kinetic equilibrium is almost identical. Assuming the NR approximation $ E_{\chi}\approx m_{\chi} $ (note that here we neglect the energy dependence on the momenta in the exponent of $\mathbf{a}_{\rm q}$, which is a more crude approximation than in Eq.~\eqref{eq:x_H_a_q}), one can write
\begin{equation} \label{eq:appx:NKE_a_and_b}
	\mathbf{a}_{\rm q}\left(x,q\right) \approx \mathbf{a}_{1}xe^{-\delta x}, \quad \mathbf{b}_{\rm q}\left(x,q\right) \approx \mathbf{a}_{1}xe^{-(\delta+1)x} \, 
\end{equation}
with
\begin{equation} \label{eq:appx:NKE_a1_deifintion}
	\mathbf{a}_{1} = \frac{6}{\pi} \sqrt{\frac{10}{g_{*}}} \frac{g_{\psi}}{g_{\chi}} \frac{m_{\psi}^{2}\alpha_{\rm NKE}m_{\rm pl}}{m_{\chi}^{2}} \frac{\sinh\left( \frac{m_{\psi}^{2}}{2m_{\chi}^{2}}\frac{q}{m_{\chi}}\bar{\beta}\right)}{q}\, .
\end{equation}
Substituting the definitions in Eq.~\eqref{eq:appx:NKE_a_and_b} into Eq.~\eqref{eq:f_chi_integrated_form} gives
\begin{equation} \label{eq:appx:NKE_f_inf}
	\bar{f}_{\chi,q,\infty} = \mathbf{a}_{1} \intop_{x_{0}}^{\infty} e^{ -\intop_{\eta}^{\infty}\mathbf{a}_{\rm NKE}\left(\xi\right)d\xi - (\delta+1)x + \ln x } d\eta \, .
\end{equation}

The saddle point $ \eta_{s} $ in this case obeys
\begin{equation} \label{eq:appx:NKE_saddle_point_value}
	e^{-\left(\delta+1\right)\eta_{s}} =\mathbf{a}_{1}^{-\frac{\delta+1}{\delta}} \left( \frac{1+\delta-\frac{1}{\eta_{s}} }{ \eta_{s}} \right)^{ \frac{\delta+1}{\delta} } \, ,
\end{equation}
and the approximations provide
\begin{multline}
	\bar{f}_{\chi,q,\infty} = \mathbf{a}_{1}^{-\frac{1}{\delta}} \left( 1+\delta-\frac{1}{\eta_{s}} \right)^{\frac{\delta+1}{\delta} }   \eta_{s}^{-\frac{1}{\delta}} e^{-\frac{1}{\delta^{2}}\left(1+\delta-\frac{1}{\eta_{s}}\right)\left(\delta+\frac{1}{\eta_{s}}\right)} \times \\ \sqrt{\frac{2\pi}{-\left(\frac{1}{\eta_{s}}\left(1+\delta-\frac{1}{\eta_{s}}\right)\left(\frac{1}{\eta_{s}}-\delta\right)-\frac{1}{\eta_{s}^{2}}\right)}} \, .
\end{multline}
Again, the $ \eta_{s} $ dependence on the mass is logarithmic and the $ \bar{f}_{\chi,q,\infty} $ dependence on the mass scale (up to logarithmic corrections) is given by the $ \mathbf{a}_{1} $ term
\begin{equation}
    \label{eq:f_chi_in_CE_approx_appendix}
	\bar{f}_{\chi,q,\infty} \sim \left( \frac{ \alpha_{\rm NKE }m_{\rm pl} }{ q } \right)^{-\frac{1}{\delta}} \, .
\end{equation}
Using Eq.~\eqref{eq:f_chi_in_CE_approx_appendix} and Eq.~\eqref{eq:Y_chi_NKE}, it is possible to estimate the DM relic abundance in the NKE equilibruim case, as done in Eq.~\eqref{eq:Y_chi_NKE_in_CE_approx} in Section~\ref{sec:NKE}.

\section{Non-integrated Boltzmann Equation for $ \chi $} \label{appx:sec:NKE_BE_derivation}

In this Appendix, we derive Eq.~\eqref{eq:f_chi_ODE} from Eq.~\eqref{eq:unintegrated_BE_for_f_chi}. The approach used here is similar to that in Refs. \cite{DAgnolo:2017dbv,Garny:2017rxs}.

Simplifying the left hand side (LHS) of Eq.~\eqref{eq:unintegrated_BE_for_f_chi} is done by changes of variables, from $ \left( t, p_{\chi} \right) $ to $ \left( x, q \right) $, with a conversion defined by $ t=t(x) $ and $ p_{\chi}= q / a(x) $.
Here, $ q $ is the comoving momentum, and in these variables the LHS is
\begin{equation}
	{\rm LHS} = \left( \frac{\partial}{\partial t} - Hp_{\chi}\frac{\partial}{\partial p_{\chi}} \right) f_{\chi} = x \frac{H}{\left(1 + \frac{1}{3} \frac{d{\rm ln}g_{*s}(T)}{d\ln T}\right)} \frac{d}{d x} \bar{f}_{\chi,q} \, ,
\end{equation}
where we ignore the $ \left(1 + \frac{1}{3} \frac{d{\rm ln}g_{*s}(T)}{d\ln T}\right) $ factor afterwords in Section \ref{sec:NKE}.

Simplifying the right hand side (RHS) of Eq.~\eqref{eq:unintegrated_BE_for_f_chi} is done in four steps. The first step is integrating over $ \boldsymbol{p}_{\phi} $, using the delta function over the 3-momentum. We obtain
\begin{multline}
	{\rm RHS} = \frac{g_{\phi}}{2 E_{\chi}} \intop \frac{1}{2E_{\phi}} d\Pi_{\psi} \left(2\pi\right) \delta\left(E_{\psi}-E_{\phi}-E_{\chi}\right) \times \\ \overline{\left|\mathcal{M}\right|^{2}}_{\psi\rightarrow\phi\chi} \left(f_{\psi}\left(E_{\psi}\right) - f_{\phi}\left(E_{\phi}\right)f_{\chi}\left(E_{\chi}\right) \right) \, ,
\end{multline}
where $ \left(\boldsymbol{p}_{\phi}\right)^{2} =\left(\boldsymbol{p}_{\psi}\right)^{2} + \left(\boldsymbol{p}_{\chi}\right)^{2} - 2\boldsymbol{p}_{\psi}\cdot\boldsymbol{p}_{\chi} $. 

The second step is integrating over $ d \Omega_{\boldsymbol{p}_{\psi}} $, where the $ z $-axis is chosen to be in the direction of $ \boldsymbol{p}_{\chi} $. Since $ \overline{\left|\mathcal{M}\right|^{2}} $ and the energies do not depend on  $ \phi_{\boldsymbol{p}_{\psi}} $, the integration over this angle is trivial and provides a factor of $ 2 \pi $.
Additionally, since $ \overline{\left|\mathcal{M}\right|^{2}} $ does not depend on $ \theta_{\boldsymbol{p}_{\psi}} $ the remaining delta function over the energy can be analytically integrated, obtaining
\begin{multline}
	{\rm RHS} = \frac{g_{\psi}g_{\phi}}{16\pi E_{\chi} } \intop_{p_{\psi}=0}^{p_{\psi}=\infty} \frac{p_{\psi}dp_{\psi}}{E_{\psi}p_{\chi}} \times \\
	 \Theta\left( 1-\left|\frac{m_{\phi}^{2}-m_{\psi}^{2}-m_{\chi}^{2}+2E_{\psi}E_{\chi}}{2p_{\psi}p_{\chi}}\right| \right) \times \\
	\overline{\left|\mathcal{M}\right|^{2}}_{\psi\rightarrow \phi \chi } \left(f_{\psi}\left(E_{\psi}\right) - f_{\phi}\left(E_{\psi}-E_{\chi}\right)f_{\chi}\left(E_{\chi}\right)\right) \, ,
\end{multline}
where $ \Theta $ is the Heaviside function.

The third step is changing the remaining integration variable from $ p_{\psi} $ to $ E_{\psi} $. This provides
\begin{multline} \label{eq:appx:unint_BE_RHS_step_3}
	{\rm RHS} =\frac{g_{\psi}g_{\phi}}{16\pi}\frac{1}{p_{\chi} E_{\chi} }\intop_{E_{\psi}=E_{\rm min}}^{E_{\psi}=E_{\rm max}}dE_{\psi}\overline{\left|\mathcal{M}\right|^{2}}_{\psi\rightarrow\phi\chi} \times \\
	\left(f_{\psi}\left(E_{\psi}\right)-f_{\phi}\left(E_{\psi}-E_{\chi}\right)f_{\chi}\left(E_{\chi}\right)\right) \, ,
\end{multline}
where (it can be shown that $ E_{\min} \ge m_{\psi} $)
\begin{equation}
	E_{\min, \max} = \delta E_{\chi} \pm \frac{ m_{\psi}^{2}\bar{\beta} p_{\chi} }{2m_{\chi}^{2}} \, .
\end{equation}

The fourth step is integrating over the energy $ E_{\psi} $. Since $ \overline{\left|\mathcal{M}\right|^{2}}_{\psi\rightarrow\phi\chi} $ does not depend on the energies of the particles and assuming both $ \psi $ and $ \phi $ are in chemical equilibrium in the Maxwell-Boltzmann distributions, Eq.~(\ref{eq:appx:unint_BE_RHS_step_3}) is simplified to
\begin{equation} \label{eq:unintegrated_BE_RHS}
	{\rm RHS} = - c\left(x = \frac{m_{\chi}}{T}, q = p_{\chi}a (x) \right) \left( f_{\chi}\left(E_{\chi}\right) - f_{\chi}^{\rm eq}\left(E_{\chi}\right) \right) \, ,
\end{equation}
with $ c \left( x , q \right) $ defined in Eq.~\eqref{eq:collition_term_NKE} in the main text.

Note that there is no need to assume a Maxwell-Boltzmann distribution for both $ \psi $ and $ \phi $. Since the final result must be of the form $ \frac{d\bar{f}_{q,\chi}}{dx} \propto - \left( \bar{f}_{q,\chi} - \bar{f}_{q,\chi}^{\rm eq} \right) $, assuming the distribution of one particle determines the proper approximation for the second, such that the BE would be consistent.

Finally, the connection between the averaged matrix element squared and the decay rate is
\begin{equation} \label{eq:appx:gamma_m_connection}
	\Gamma = \frac{1}{2m_{\psi}}g_{\chi}g_{\phi} \overline{\left|\mathcal{M}\right|^{2}}_{\psi \rightarrow \chi \phi} \frac{\bar{\beta}}{8\pi} \, ,
\end{equation}
and so in the language of the parameterization of Eq.~ \eqref{eq:gamma_parameterization}, we have
\begin{equation}
	\overline{\left|\mathcal{M}\right|^{2}}_{\psi \rightarrow \chi \phi} = \frac{16\pi m_{\psi}^{2}\alpha_{\rm decay}}{g_{\chi}g_{\phi}} \, .
\end{equation}

\bibliography{biblio}{}

\end{document}